\documentclass{article}
\usepackage{authblk}
\usepackage[utf8]{inputenc}
\usepackage[margin=2cm]{geometry}
\usepackage{amsmath}
\usepackage{amssymb}
\usepackage{graphicx}
\usepackage[group-separator={,}]{siunitx}
\usepackage{subfig}
\usepackage{xcolor}
\usepackage{cite}

\title{The Polycentric Dynamics of Melbourne and Sydney: Suburb attractiveness divides a city at the home ownership level}

\author[1,2]{Emanuele Crosato}
\author[1]{Mikhail Prokopenko}
\author[1]{Michael S. Harr\'e}

\affil[1]{Complex Systems Research Group, Faculty of Engineering, The University of Sydney, Sydney, Australia}
\affil[2]{School of Physics and EMBL Australia node in Single Molecule Science, School of Medicine, UNSW, Sydney, Australia}

\date{}

\begin{document}
 
\maketitle

\begin{abstract}
Urban dynamics in large metropolitan areas result from complex interactions across social, economic, and political factors, including population distribution, flows of wealth, and infrastructure requirements. We develop a Census-calibrated model of urban dynamics for the Greater Sydney and Melbourne areas for 2011 and 2016, highlighting the evolution of population distributions and the housing market structure in these two cities in terms of their mortgage and rent distributions. We show that there is a tendency to homophily between renters and mortgage holders: renters tend to cluster nearer commercial centres, whereas mortgagors tend to populate the outskirts of these centres. We also identify a critical threshold at which the long-term evolution of these two cities will bifurcate between a `sprawling' and a `polycentric' configuration, showing that both cities lie on the polycentric side of the critical point in the long-run. Importantly, there is a divergence of these centric tendencies between the renters and mortgage holders. The polycentric patterns characterising the mortgagors are focused around commercial centres, and we show that the emergent housing patterns follow the major transport routes through the cities.
\end{abstract}


\section{Introduction}

Cities are multi-dimensional dynamical systems with a complex array of dynamic interactions and pattern-forming behaviours~\cite{Fujita1982,bettencourt2007growth,wilson2008boltzmann,bettencourt2013origins,Barthelemy2016book,Osawa2017,Ellam2018,Barthelemy2019}. While there have been significant advances in understanding complex micro- and meso-scale urban dynamics and interactions, their relationships with dispersed infrastructure, transport and service networks, as well as their dependencies on a spatial distribution of financial and industrial centres, remain poorly understood. In particular, a dynamic evolution of resettlement patterns and intra-city migration flows, constrained by urban planning and financial, real-estate, and social pressures, is hard to quantify and predict, even with abundance of historic and contemporary data sources. There is a multiplicity of hidden variables and exogenous factors affecting urban evolution, including population density, hierarchical urban structures, low-density settlement and urban sprawl, polycentric transitions, human mobility, and vulnerability to climatic stress, pandemics and other crises~\cite{Simini2012,Louf2013,Bouchaud2013,dearden2015explorations,Arcaute2016,Volpati2018,Penny2018,Barbosa2018,Zachreson2018,Fletcher2019,Wu2019,slavko2019dynamic,harding2020population}.  
And yet, these are critical and inter-dependent issues that we need to understand if we are to plan for larger, safer and more affluent cities. Ideally, the future cities may be either monocentric or polycentric, combine high and low densities, accommodate renters, investors and home owners, balance ecological and industrial priorities, while being resilient and adaptive to challenges of the modern world~\cite{crosato2018critical,Barthelemy2013Planning,slavko2020city}.

The most granular scale for which we usually have access to data is the household level, and at this level we can model micro-economic decisions based on each household's budgetary constraints~\cite{baptista2016macroprudential,axtell2014agent,glavatskiy2020explaining}. For example, two similar households desiring to live in the same suburb and only differing in their level of stock of wealth (assets that can, for example, be readily converted to a deposit on a mortgage) might have two different outcomes: one might choose to take a mortgage and settle within the suburb while the other might remain in the suburb but continue to rent. In the first case, if wage growth is good, the value of the mortgage repayments will drop in real terms, and therefore, the real cost of housing services to this household decreases over time. In the case of the second household, if house prices appreciate at roughly the same rate as wages and the rental price reflects the market price of the property (rather than the purchase price), the real cost of housing services (rent) will remain roughly proportional to income over time. This can lead to an increasing economic disparity between home owners and home renters, as is the case in Australia~\cite{burbidge2000capital} and elsewhere~\cite{frick2003imputed}. This is observed particularly if the two effects combine with the ability to secure a new mortgage using the mark-to-market value of a house for private, rather than institutional, investing~\cite{lee2017examination}: strong wage growth and strong housing asset appreciation make it easier for current home owners to purchase a second (or more) rental property, increasing demand for houses and further pulling house prices away from the reach of currently renting households.

Compounding these issues are the long term effects on social mobility and economic inequality that home ownership can entrench. Inter-generational wealth transfers through deceased estates provides a direct mechanism for the first home buyers whose parents owned a home to be advantaged over those whose parents were not home owners~\cite{engelhardt1998intergenerational,cigdem2017intergenerational}. Similarly home ownership for offspring that can be secured by the home owning parents can also provide their children with an advantage not available to those whose parents are not home owners. It has been noted~\cite{coulter2018parental} that these mechanisms can further entrench renting producing an inter-generational ``economically marginalised'' group of renters. On the other hand, the opportunity to accumulate wealth in the form of housing earlier in life provides the benefits of a decreasing real cost of housing services mentioned above: the earlier a house is acquired the greater the cumulative effect as the initial cost of a mortgage is depreciated. 

These long term socio-economic issues need to be connected with current population distributions and individuals' physical mobility within a city, for example via population growth-diffusion processes~\cite{pred1975diffusion,li2003simulating,batty2006hierarchy,pumain2006evolutionary,wu2017city}. From this perspective, diffusion as a process of urban expansion plays an important role in the evolution of cities. Earlier studies have found that using existing population distributions is predictive of current mobility patterns~\cite{yan2014universal}, that income $i$ and commuting distances $r$ are non-linearly related by the power law $r \propto i^{\beta}$ with $\beta \in [0.5,0.9]$~\cite{carra2016modelling}, and that transport patterns have typical behavioural motifs~\cite{schneider2013unravelling}. These studies provide insights into the current state of cities but not their future evolution, their relationship to an ``optimal'' configuration, or the consequences for long-term policy making, although progress has been made in this direction~\cite{muneepeerakul2012critical,suen2019travel}.

In order to understand the short term dynamics for the greater Sydney and Melbourne areas (Australia) that contribute to long term socio-demographic and economic trends we use a dynamical systems modelling approach first introduced by Wilson~\cite{wilson2008boltzmann}. This approach uses the maximum entropy principle (MaxEnt) to minimise a loss function given certain constraints. In our approach, the loss function is the (negative) entropy and the constraints are travel costs, suburb attractiveness based on residential type (owner or renter), and the size of the residential population in each suburb. This provides a least biased estimate of the (constrained) population distribution of renters and home owners. This static distribution is then coupled to a set of Lotka-Volterra dynamical equations that redistributes the population based on the suburbs ``competing'' for residents based on the perceived attractiveness to renters or home owners, see~\cite{dearden2015explorations,dearden2012relationship}. The data used is obtained from the Australian census records, and provides a suburb by suburb dynamical representation of the Australian housing market in Melbourne and Sydney for 2011 and 2016.

This approach can be compared with the Population Weighted Opportunity model~\cite{yan2014universal} that produces a parameter free prediction of mobility patterns based on extant population distributions. In that model ``attractiveness'' (a proxy for opportunities in a local area of the city) is a linear function of a suburb's population size. It is assumed that the attractiveness of the destination is the product of local opportunities and the inverse of the total population in a region surrounding the destination. Aside from the attractiveness function itself, the main difference between that approach and ours (similarly to the discussion above) is in that we are inferring the long term dynamical evolution of a population based on income, housing costs, population, and travel costs, rather than using the current population to infer mobility.

In this study, we investigate urban dynamics of Melbourne and Sydney, and report three main results. Firstly, we quantify settlement preferences for renters and mortgage holders and identify a tendency to homophily. This tendency is manifested in the choices of renters preferring to live in proximity to commercial centres (including central business districts, CBDs), while mortgagors are found to predominantly populate the outskirts of these centres and suburbia in general. Secondly, we pinpoint a critical threshold at which the long term evolution of Melbourne and Sydney is shown to bifurcate between a ``sprawling'' configuration and a ``polycentric'' configuration. Both cities are projected to lie on the polycentric side of the critical point in the long-run, with commercial centres shaping the polycentric patterns. Finally, we describe an housing pattern in Sydney, and relate this structure to major transport routes through the city.


\section{Model and numerical simulation}


\subsection{Description of the main quantities}

In this study we consider the \emph{working} population of urban agglomerations that is either paying mortgage or renting a property---any other category (e.g. children, unemployed people, people who already own a house, etc.) is excluded.
The considered urban agglomerations cover geographical areas that are divided into $N_\text{res}$ \emph{residential} suburbs, as well as into $N_\text{emp}$ \emph{employment} suburbs, whose quantity and shapes can be different from the ones of the residential suburbs.
Each individual works in an employment suburb $i$ and lives in a residential suburb $j$ and $i$ might equal $j$, hence, we describe the number of people who daily commute between employment and residence suburbs with  the matrix $T_{ij}$. Let us also denote $P_j=\sum_i T_{ij}$ the (working) population living in suburb $j$, $E_i=\sum_j T_{ij}$ the population employed in suburb $i$, and $P=\sum_{i,j} T_{ij}$ the total population.

The commuting matrix can be divided into a renters part $T^\text{R}_{ij}$ and a mortgagors part $T^\text{M}_{ij}$, i.e.
\begin{equation}
T_{ij}=T^\text{R}_{ij}+T^\text{M}_{ij}.    
\end{equation}
Let $P^\text{R}_j = \sum_i T^\text{R}_{ij}$ and $P^\text{M}_j = \sum_i T^\text{M}_{ij}$ denote, respectively, the number of people who are renting a property in $j$ and the number of people who are paying a mortgage in $j$.
Similarly, we divide the population working in suburb $i$ into $E^\text{R}_i=\sum_j T^\text{R}_{ij}$ and $E^\text{M}_i = \sum_j T^\text{M}_{ij}$.

Every commute from a residential suburb $j$ to an employment suburb $i$ has an associated cost $C_{ij}$ (e.g. expenses, time, hazard, etc.). Additionally, residential suburbs are characterised by an average weekly rent $R_j$ and an average weekly mortgage $M_j$, while employment suburbs are characterised by an average weekly income $I_i$. For example, a person working in $i$ and renting in $j$ would receive a weekly income $I_i$, commute to work at a cost $C_{ij}$ and pay a weekly rent $R_j$, while a person working in $i$ and having a mortgage in $j$ would receives a weekly income $I_i$ and commutes at a cost $C_{ij}$, but pays a weekly mortgage of $M_j$ instead. Importantly, in our model some residential suburbs are more attractive than other, depending on factors such as rent, mortgage and population. The preference to live in a suburb rather than another also depends on whether the person is buying or renting. Specifically, we define the `attractiveness' $A^\text{R}_j$ for renters and $A^\text{M}_j$ for mortgagors as follows:
\begin{equation}
\label{eq:attr-rent}
A^\text{R}_j=\log(1+(\hat{R}-R_j)(\hat{M}-M_j)P_j)
\end{equation}
and
\begin{equation}
\label{eq:attr-mort}
A^\text{M}_j=\log(1+(\hat{R}-R_j)(\hat{M}-M_j)P^\text{M}_j) ,
\end{equation}
where $\hat{R}=2\max R_j$ and $\hat{M}=2\max M_j$ are thresholds for the maximum values of rent and mortgage respectively (in the case of Sydney in year 2016, for example, $\hat{R}=\num{1758}$ and $\hat{M}=\num{1943}$).
The statistical validation for these definitions of attractiveness are provided in Appendix~\ref{sec:attractiveness-functions}.


\subsection{The BLV model}

The initial values of $E_i^\text{R}$, $E_i^\text{M}$, $C_{ij}$, $I_i$, $R_j$, $M_j$ (and therefore $A_j^*$) are known\footnote{Australian Bureau of Statistics: https://www.abs.gov.au/}.
We also assume that the employment place and transport costs are fixed, i.e. the quantities $E^\text{R}_i$, $E^\text{M}_i$, and $C_{i,j}$ do not change. We are then interested in predicting the commuting matrix $T_{ij}$ as the population redistributes among the residential suburbs as a tradeoff between the commuting cost $C_{ij}$ and residential suburbs' attractiveness $A_j$.
Crucially, the redistribution of the population $T_{ij}$ causes the income flow $Y_j=\sum_i T_{ij}I_i$ to change, which in turn affects the evolution of the weekly rent and mortgage and, consequently, the attractiveness.
Thus, a loop is formed: as rent and mortgage change over time because of the population redistribution, the distribution of population changes because of the variations in rent and mortgage.
We are interested in observing towards what configuration the commuting matrix $T_{ij}$ evolve as a consequence of this loop.
Importantly, we assume the existence of a \emph{slow} and a \emph{fast} dynamics: the commuting matrix changes much quicker than rent and mortgage.
This means that, in our model, the population is instantaneously redistributed across the suburbs when the suburbs' attractiveness changes.

In this study the urban dynamics of Greater Sydney and Greater Melbourne are modelled using an adaptation of the Boltzmann-Lotka-Volterra (BLV) model~\cite{wilson2008boltzmann}, calibrated to Census data and geospatial information.
Such model is characterised by two components: a maximum entropy (Boltzmann) component, describing the fast dynamics of the population redistribution, and a dynamical (Lotka-Volterra) component, describing the slow dynamics of rent and mortgage change.


\subsubsection{The maximum entropy component}

This component consists of calculating, for both renters and mortgagors, the least biased commuting matrices that are consistent with some constraints.
Specifically, we calculate the $T^\text{*}_{ij}$ that maximises the entropy
\begin{equation}
\label{eq:maxent}
H(T^\text{*}_{ij}) = -\sum_{i=1}^{N_\text{emp}}\sum_{j=1}^{N_\text{res}} T^\text{*}_{ij}\log T^\text{*}_{ij} ,
\end{equation}
subject to the constraints
\begin{equation}
\label{eq:constraint_1}
\sum_{j=1}^{N_\text{res}} T^\text{*}_{ij} = E^\text{*}_i ,
\end{equation}
\begin{equation}
\label{eq:constraint_2}
\sum_{i=1}^{N_\text{emp}}\sum_{j=1}^{N_\text{res}} T^\text{*}_{ij}C_{ij} = C^\text{*}_\text{tot} .
\end{equation}
\begin{equation}
\label{eq:constraint_3}
\sum_{i=1}^{N_\text{emp}}\sum_{j=1}^{N_\text{res}} T^\text{*}_{ij}A^\text{*}_{j} = A^\text{*}_\text{tot} ,
\end{equation}
where `*' is either `R', indicating renters, or `M', indicating mortgagors, $C^\text{*}_\text{tot}$ is the total cost of travelling for the entire population and $A^\text{*}_\text{tot}$ is the total attractiveness to suburbs.
The constraint in Eq.~\eqref{eq:constraint_1} fixes total amount of workers in each employment suburb.
The constraint in Eq.~\eqref{eq:constraint_2} sets the total cost of travelling for the entire population and, similarly, the constraint in Eq.~\eqref{eq:constraint_3} sets the total attractiveness for the entire population.

The solution to this problem is
\begin{equation}
\label{eq:boltzmann-solution}
T^\text{*}_{ij} = \frac{E^\text{*}_i e^{\alpha^\text{*} A^\text{*}_j - \gamma^\text{*} C_{ij}}}{Z^\text{*}_i} ,
\end{equation}
where
\begin{equation}
    Z^\text{*}_i = \sum_{k=1}^{N_\text{rev}}e^{\alpha^\text{*} A^\text{*}_k - \gamma^\text{*} C_{ik}}
\end{equation}
is analogous to a partition function. The coefficients $\alpha$ and $\gamma$ are the Lagrangian multipliers corresponding to the constraints in Eqs.~(\ref{eq:constraint_2}) and (\ref{eq:constraint_3}).
They represent, respectively, the \emph{social disposition}, i.e. to what extent people are affected by suburbs' attractiveness, and the \emph{impedance to travel}, i.e. to what extent people are affected by commuting travels, respectively.
The values of $\alpha$ and $\gamma$ largely affect the behaviour of the system and their influence is studied in this work.
The parameters $\alpha$ and $\gamma$ can also be used for calibrating the model to Census data, as explained in the Sec.~\ref{sec:cal-sim}.


\subsubsection{The dynamical component}

Suburbs' average weekly rent and mortgage vary in time $t$ according to the following dynamics:
\begin{equation}
\label{eq:LVrent}
\frac{dR_{j}}{dt} = \epsilon(k^\text{R}Y^{\text{R}}_{j} - P^\text{R}_jR_j) ,
\end{equation}
\begin{equation}
\label{eq:LVmort}
\frac{dM_{j}}{dt} = \epsilon(k^\text{M}Y^{\text{M}}_{j} - P^\text{M}_jM_j) ,
\end{equation}
where $Y_j^\text{*}=\sum_i T^\text{*}_{ij}I_i$ and $\epsilon$, $k^\text{R}$ and $k^\text{M}$ are constants.
Specifically, $\epsilon$ determines the size of the change, $k^\text{R}$
represents the fraction of peoples' salary that is spent in rent and $k^\text{M}$
represents the fraction of peoples' salary that is spent in mortgage (e.g., $k^\text{R}=0.3586$ and $k^\text{M}=0.4717$ for Sydney in 2016).

\begin{figure}[t!]
    \centering
    \includegraphics[width=\columnwidth]{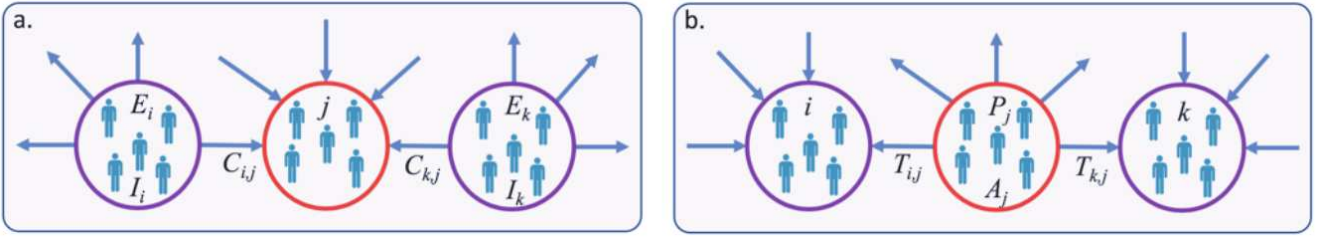}
    \caption{Purple circles are the suburbs people work in, the red circle is a suburb people live in. People can work in the same suburb they reside in. (a) Static values are the population of workers in each suburb $E_i$ and $E_k$, sources of income for suburb $j$: $I_i$ and $I_k$, and the travel costs $C_{ij}$ and $C_{kj}$. (b) Dynamical values are the attractiveness of a suburb $A_j$, the residential population $P_j$,and the commuter network $T_{ij}$.}
    \label{fig:local_network_structure}
\end{figure}

The maximum entropy commuting matrix $T_{ij}$ is recomputed at every change of rent and mortgage.
Therefore, running the BLV model consists of repeating the following two steps in a loop:
\begin{enumerate}
\item calculate the maximum entropy $T^\text{*}_{ij}$ using Eq.~\eqref{eq:boltzmann-solution} using the most recent values of $A^\text{*}$;
\item update $R_j$ and $M_j$ according to Eqs.~(\ref{eq:LVrent}-\ref{eq:LVmort}) and, consequently, update the attractiveness $A^\text{*}$.
\end{enumerate}
until the change in $T^*_{ij}$ between two consecutive iterations becomes negligible, i.e. less than a small constant used to test approximate convergence. See Figure~\ref{fig:local_network_structure} for the relationship between the static and the dynamical variables as a local network structure of the suburbs. 

\subsection{Calibration and simulation}
\label{sec:cal-sim}

We consider the cases of Greater Sydney and Greater Melbourne in years 2011 and 2016, for which we have Census and a detailed geospatial description of residential and employment suburbs.

The Census data provides the commuting matrix ${\cal T}_{ij}$, and its division into ${\cal T}^\text{R}_{ij}$ and ${\cal T}^\text{M}_{ij}$ can also be derived---note that a calligraphic symbol is used in order to distinguish this Census commuting matrix ${\cal T}_{ij}$, utilised for initialisation and for calibration, from the commuting matrix $T_{ij}$ predicted by the model.
From ${\cal T}^\text{R}_{ij}$ and ${\cal T}^\text{M}_{ij}$ we can determine $E^\text{R}_i$ and $E^\text{M}_i$ which, we remind, don't change.
The Census also provides the initial values for $R_j$, $M_j$, from which the initial attractiveness $A_j$ can be calculated using Eqs.~\eqref{eq:attr-rent} and~\eqref{eq:attr-mort}.
Finally, the cost of commuting $C_{ij}$ is estimated as the Euclidean distance between the centroids of suburbs $i$ and $j$.
A more precise estimation of $C_{ij}$ would be possible if given access to trips data (such as time and cost) which includes all travelling modes (i.e. car, train, ferry, etc.).
However, if not all travelling modes are considered, the estimation will be biased (think for example of not considering the ferry in for a city like Sydney which is built around a harbor) and thus, for this exploratory study, we used a simple approximation (adopted in an earlier study~\cite{crosato2018critical}).

\begin{table}[b]
\centering
\begin{tabular}{|c|c|c|c|c|}
\hline
City & $\hat{\alpha}^\text{R}$ & $\hat{\gamma}^\text{R}$ & $\hat{\alpha}^\text{M}$ & $\hat{\gamma}^\text{M}$\\
\hline
Sydney 2016 & 1.0 & 0.022 & 1.4 & 0.020\\
Sydney 2011 & 1.1 & 0.025 & 1.5 & 0.020\\
Melbourne 2016 & 1.1 & 0.020 & 1.5 & 0.017\\
Melbourne 2011 & 1.1 & 0.022 & 1.5 & 0.018\\
\hline
\end{tabular}
\caption{Summary of the values $\hat{\alpha}^\text{R}$, $\hat{\alpha}^\text{M}$, $\hat{\gamma}^\text{R}$ and $\hat{\gamma}^\text{M}$ that best match the Census data.}
\label{tab:opt-val}
\end{table}

We this data, we can now calculate the maximum entropy $T^\text{R(start)}_{ij}$ and $T^\text{M(start)}_{ij}$ using Eq.~\eqref{eq:boltzmann-solution}.
Here the label `start' is used in order to highlight that these are the first calculated values of the commuting matrices, i.e. before the dynamics take place.
Thus, the calibration of the model therefore consists in calculating $T^\text{R(start)}_{ij}$ and $T^\text{M(start)}_{ij}$ for several combinations of values of the social dispositions $\alpha^\text{R}$ and $\alpha^\text{M}$ and the travel impedances $\gamma^\text{R}$ and $\gamma^\text{M}$, and contrasting them with the Census commuting matrices ${\cal T}^\text{R}_{ij}$ and ${\cal T}^\text{M}_{ij}$ to find the best matching combination (see details in Appendix~\ref{sec:calibration}).
We stress that such calibration is carried out before the dynamics, since our purpose is to investigate to what configurations the urban dynamics of Greater Sydney or Melbourne are relaxing towards, given the initial conditions.
The best matching values of $\hat{\alpha}^\text{R}$, $\hat{\gamma}^\text{R}$, $\hat{\alpha}^\text{M}$ and $\hat{\gamma}^\text{M}$ are summarised in Table~\ref{tab:opt-val}.
These values are quite similar across years and, interestingly, also between the two different cities.

The constants $k^\text{R}$ and $k^\text{M}$ have also been calibrated setting their values to
\begin{equation}
    k^\text{R} = \frac{\sum_{j=1}^{N_\text{res}} P^\text{R}_jR_j}{\sum_{i=1}^{N_\text{emp}} Y^\text{R}_i}
\end{equation}
and
\begin{equation}
    k^\text{M} = \frac{\sum_{j=1}^{N_\text{res}} P^\text{M}_jM_j}{\sum_{i=1}^{N_\text{emp}} Y^\text{M}_i} . 
\end{equation}
Hence, the constant $k^\text{R}$ (or $k^\text{M}$) is set to the ratio between the total amount of money spent in rent (or mortgage) by the entire population and the total amount of money earned by the entire population.
The constants are set at the beginning, based on the population, rent, mortgage and income obtained from the Census.

Once the best matching values $\hat{\alpha}^\text{R}$, $\hat{\alpha}^\text{M}$, $\hat{\gamma}^\text{R}$ and $\hat{\gamma}^\text{M}$ are identified, we explore the dynamics of the system at $\hat{\gamma}^\text{R}$ and $\hat{\gamma}^\text{M}$ and within an interval around $\hat{\alpha}^\text{R}$ and $\hat{\alpha}^\text{M}$.
More specifically, we introduce a control parameter, $\rho$, which function as a scaler for the social disposition of both renters and mortgagors, so that
\begin{equation}
    T^\text{*}_{ij} = \frac{E^\text{*}_i e^{\rho\hat{\alpha}^\text{*} A^\text{*}_j - \hat{\gamma}^\text{*} C_{ij}}}{Z^\text{*}_i}
\end{equation}
where, once again, `*' is either `R', indicating renters, or `M', indicating mortgagors.
We explore an interval of $\rho$ between $0$ and $1.5$ at small steps (note that $\rho=1$ corresponds to the best matching values of social disposition) by running the BLV model for each value of $\rho$ until changes in $T_{ij}$ become negligible.


\section{Results}


\subsection{Two urban configurations}

\begin{figure}[b!]
    \centering
    \subfloat[]{
        \includegraphics[width=0.49\columnwidth,trim={1.1cm 0.8cm 0.8cm 0.3cm},clip]{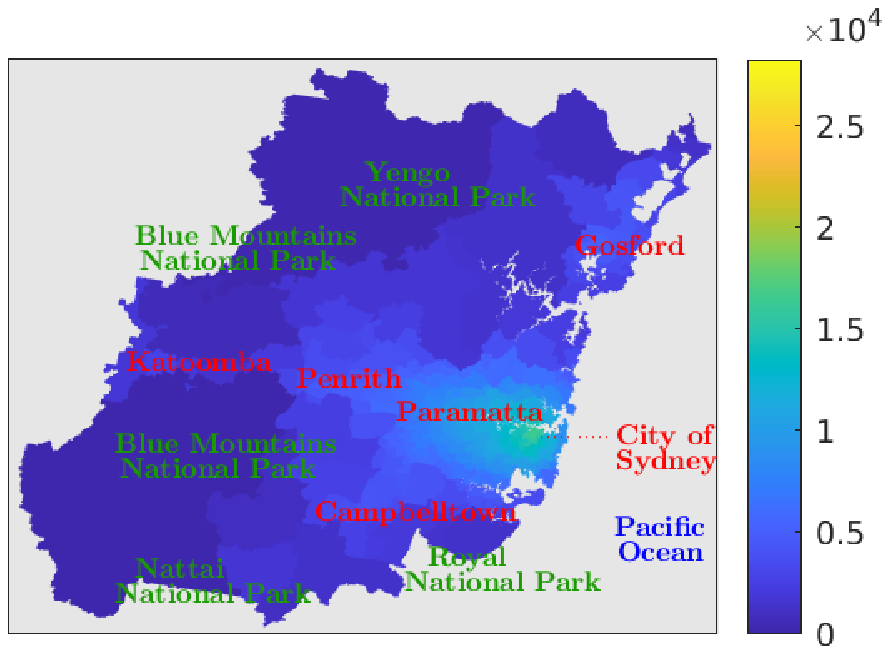}
        \label{fig:sydney-maps-2016-sprawl}}
    \subfloat[]{
        \includegraphics[width=0.49\columnwidth,trim={1.1cm 0.8cm 0.8cm 0.3cm},clip]{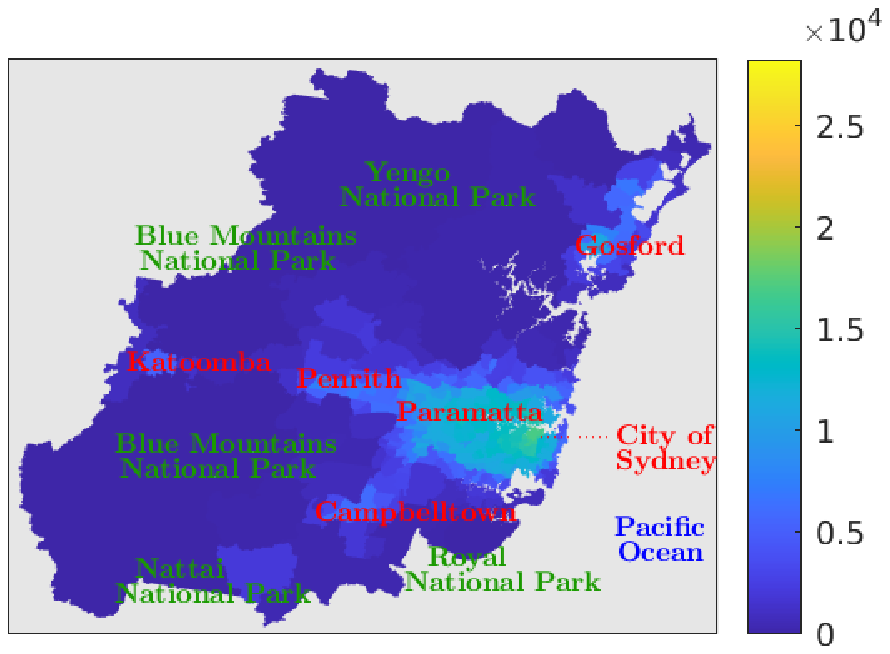}
        \label{fig:sydney-maps-2016-critical}}\\
    \subfloat[]{
        \includegraphics[width=0.49\columnwidth,trim={1.1cm 0.8cm 0.8cm 0.3cm},clip]{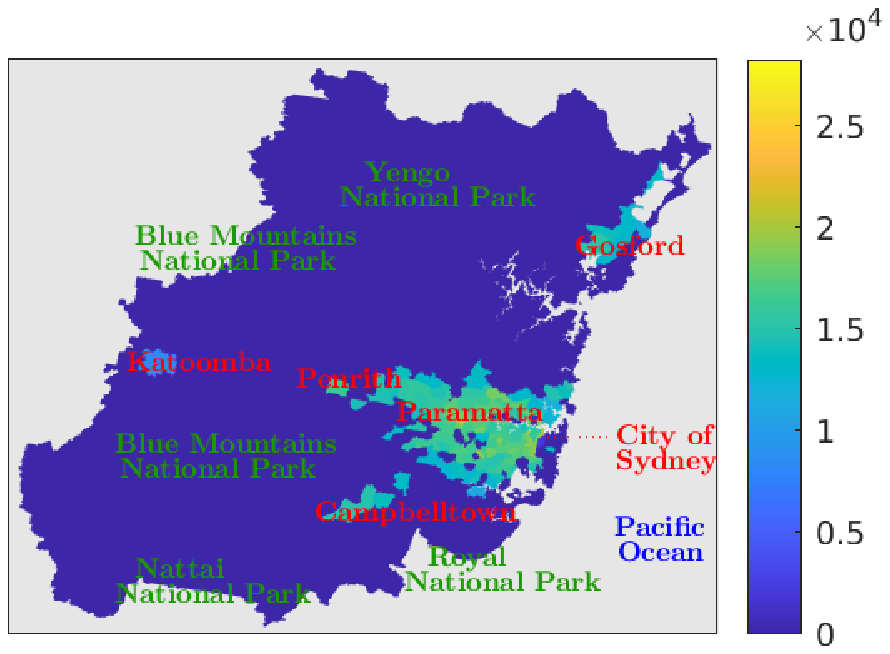}
        \label{fig:sydney-maps-2016-poly}}
    \subfloat[]{
        \includegraphics[width=0.49\columnwidth]{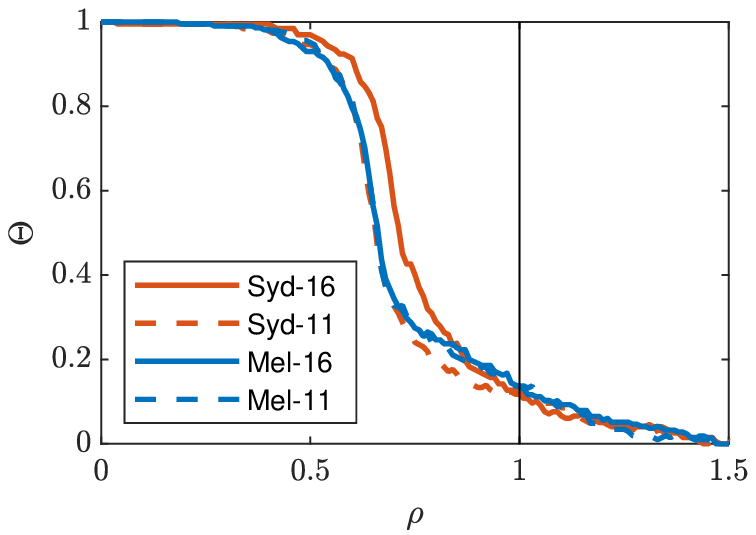}
        \label{fig:trans-comp}}%
    \caption{(a-c) Distribution of the population $P_j$ over the residential suburbs $j$ of Greater Sydney (2016), after running the BLV dynamics using (a) $\rho=0.2$, (b) $\rho=0.6$ and (c) $\rho=1$
    (d) Fraction of residential suburbs with more than $300$ inhabitants remaining at the end of the run, over values of $\rho$ from $0$ to $1.5$. The vertical bar at $\rho=1$ indicates the best match with Census data. The graph compares Greater Sydney and Greater Melbourne in the years 2011 and 2016. (e-f) the attractiveness of each suburb for renters (e) and mortgagors (f) for $\rho =1$.}
    \label{fig:sydney-maps-2016}
\end{figure}

For all four considered cases (i.e., Greater Sydney and Greater Melbourne in years 2011 and 2016) we observe a sudden change in the final distribution of the population, as we vary the control parameter $\rho$.
This phenomenon is illustrated in Fig.~\ref{fig:sydney-maps-2016} for Sydney (2016), while the other three cases are provided as Supplemental Material (SM).
The figure shows a map of Sydney displaying the final population $P_j$ of all residential suburb $j$ obtained with three different values of $\rho$.

With small values of $\rho$, the population distributes almost homogeneously across the geographical region, with a slightly higher concentration around the Central Business District (CBD)---see Fig.\ref{fig:sydney-maps-2016-sprawl}.
This configuration is sometimes referred as ``sprawling''~\cite{crosato2018critical}.
Increasing the value of $\rho$ does not largely affect the distribution of the population, until a critical value is approached (see Fig.\ref{fig:sydney-maps-2016-critical}) at which point the distribution of the population abruptly moves towards a polycentric configuration, where most of the population starts to distribute around a few more highly populated centres.
Increasing $\rho$ even further does not produce different urban configurations, but rather only reinforces the polycentric character of the population distribution (see Fig.\ref{fig:sydney-maps-2016-poly}).

\begin{figure}[b!]
    \centering
    \subfloat[]{
        \includegraphics[width=0.49\columnwidth,trim={1.1cm 1.2cm 0.7cm 0.6cm},clip]{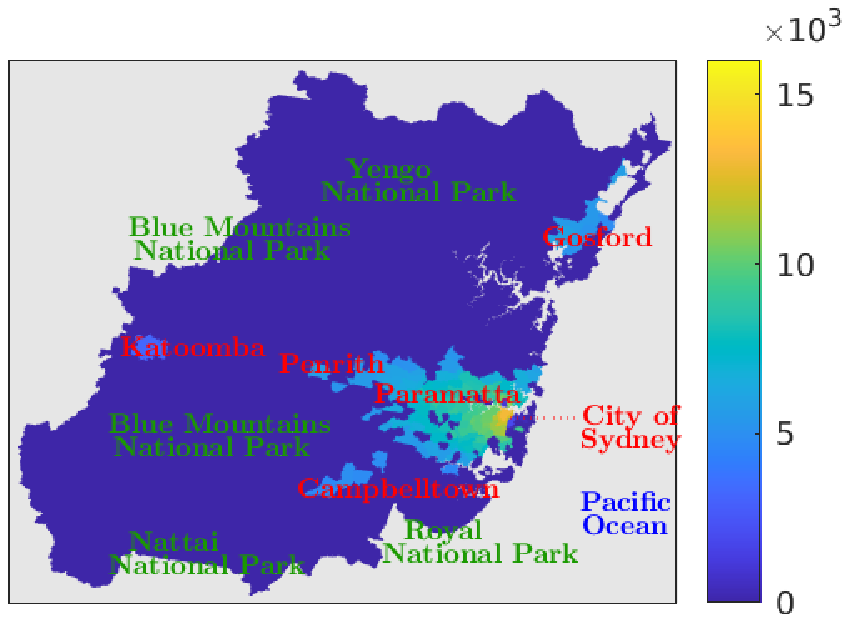}
        \label{fig:sydney-renters-2016}}
    \subfloat[]{
        \includegraphics[width=0.49\columnwidth,trim={1.1cm 0.8cm 0.8cm 0.3cm},clip]{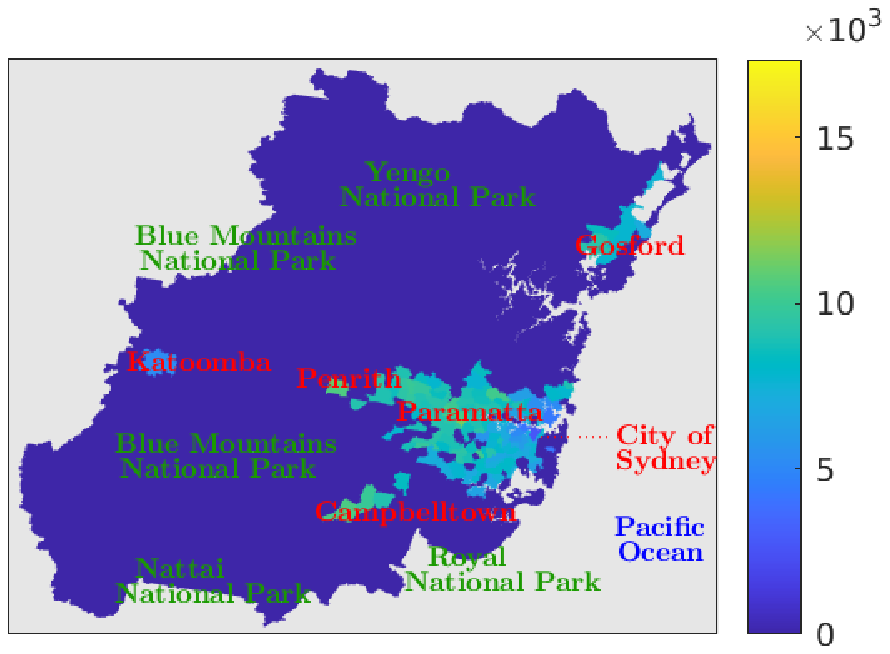}
        \label{fig:sydney-mortgagors-2016}}\\
    \subfloat[]{
        \includegraphics[width=0.6\columnwidth,trim={0.0cm 0.0cm 0.0cm 0.0cm},clip]{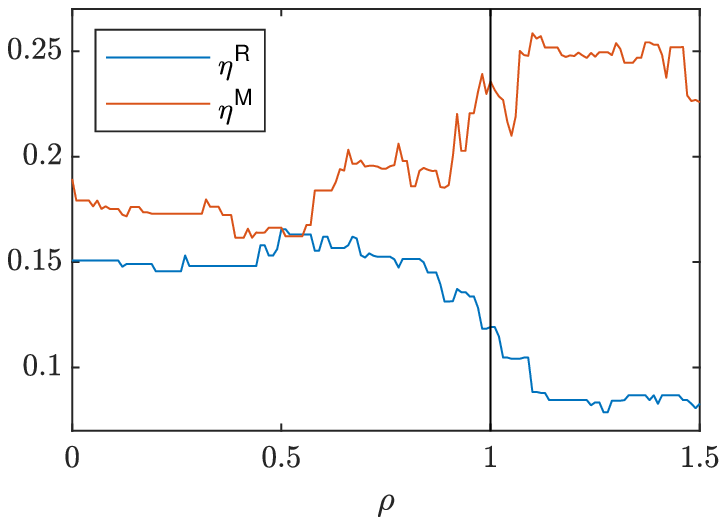}
        \label{fig:sydney-spreading-2016}}
    \caption{(a-b) Population distribution of renters (a) and morgagors (b) for Sydney 2016 at the end of simulation, $\rho = 1$. (c) The spreading-index~\cite{louail2014mobile} for Sydney 2016 calculated at the end of the simulation with different values of $\rho$.}
    \label{fig:sydney-renters-mortgagors-2016}
\end{figure}

This sharp transition can be analysed through the introduction of an order parameter, $\Theta$, which quantifies the fraction of suburbs that have a population larger than a certain threshold.
For example, in Fig.~\ref{fig:trans-comp} $\Theta$ is mapped over $\rho$ choosing a threshold of $300$ people.
We can observe that for small values of $\rho$ all suburbs have more that $300$ people, while for high values of $\rho$ only about $30$\% of them (corresponding to the emerging major centres) have a population that exceeds this threshold.
The transition is captured by a sudden drop in the order parameter $\Theta$ at approximately $\rho = 0.6$ (or $\rho = 0.65$ in the case of Sydney 2016).
Importantly, we note that in each of the four cases (i.e. Greater Sydney and Greater Melbourne in 2011 and 2016) the critical $\rho$ is below the value 1, indicating that the population of both cities would settle, in the long run, on a polycentric configuration as these cities evolved according to the BLV dynamics considered here.

It is also interesting to note that the critical value of $\rho$ is very similar for Melbourne in both Census years and Sydney in 2011, while for Sydney 2016 it is slightly higher. The results suggest that Sydney and Melbourne are quite comparable in terms of urban dynamics, however, in the most recent years Sydney's relaxation configuration has moved towards the sprawling configuration, while still residing within the polycentric configuration. 
What drives this polycentricity are the competing industrial and urban centres that have developed as Sydney has expanded. To the north, west, and south of the CBD industrial centres offering employment opportunities have developed on land that was historically cheap, pulling workers away from the financial centre of the CBD, making it more attractive to rent or purchase property closer to these new industrial centres. This can be seen in the population spikes in Gosford (north), Campbelltown (south), and Paramatta and Penrith (west). The recent changes in Sydney from 2011 to 2016 shown in these plots suggest that these population centres are becoming less influential in deciding where people live. In particular, as the costs of housing in an urban centre rise, it may be preferable to balance the higher travel costs but lower property values by living further away from an urban centre, but as there are multiple urban centres near one another, as people move outwards from one centre they move towards another, thereby diffusing the Polycentricity of the urban landscape over time, creating a more sprawling urban environment. We further split this analysis out into renters and homeowners next.


\subsection{Distribution of renters and mortgagors}

In the previous section, we have described how changes in the parameter $\rho$ affect the distribution of the whole population, including both renters and mortgagors.
Interestingly, however, we notice that renters and mortgagors are characterised by very different patterns of distribution across the urban areas. In fact, it is in general observed that the renting population is more concentrated around the CBD and less concentrated in the peripheral suburbs, while the mortgagor population exhibits the opposite trend, both populations exhibiting the type of homophilic tendencies typically observed social situations~\cite{mcpherson2001birds}. This phenomenon is observed in both the sprawling and the polycentric configurations, although in the latter it is more notable (e.g., Fig.\ref{fig:sydney-renters-mortgagors-2016}).

These interesting spatial divisions cannot be predicted directly from the expressions of the attractiveness nor the constraints and dynamics of the model. Instead they are a consequence of the dynamic interactions between the cost of housing, population, and the location of jobs.
Recall that $E_i$ and $C_{ij}$, the employment  count in each suburb and the transport costs of travelling between suburbs, are fixed values, whereas $P_j$ and $T_{ij}$ are the residential population and commuting population flows that are derived iteratively via the MaxEnt and BLV dynamics. Also recall that a key driver of these dynamics is the attractiveness $A_j$ of each suburb.  This means that $T_{ij}$ is the network along which there is a flow of household income from a source node $i$ to destination node $j$ with a cost of $C_{ij}$. Partitioning these flows according to renter and mortgagor populations produce quite different population dispersion patterns as shown in Figure~\ref{fig:sydney-renters-mortgagors-2016}. What we see after simulation is an aggregation of populations around the known commercial centres, but within these aggregations patterns of population microstructures can be seen as the renters move closer to the commercial centre and the mortgagors move to the fringes; i.e. there is an economic homophily effect as like is attracted to like in terms of housing arrangements.

In order to quantify polycentricity of urban agglomerations, we use the spreading-index~\cite{louail2014mobile}, which is expected to be close to 0 for perfectly monocentric agglomerations, and approach 1 for polycentric agglomerations.
When applied to our projections of Sydney's population (based on 2016 data), the spreading index reveals a clear divergence between the populations of renters and mortgagors (see Fig.~\ref{fig:sydney-spreading-2016}).
For low values of $\rho$, corresponding to the sprawling phase, the spreading index is quite stable around $0.15$ for renters and $0.17$ for mortgagors.
These values of $\rho$ correspond to an urban configuration that is neither monocentric nor polycentric, but rather homogeneously distributed over the whole geographical area.
However, as we increase $\rho$ and drive the population over the transition, into a ``centric'' configuration, the spreading index increases to approximately $0.25$ for mortgagors, indicating growing polycentricity, and decreases to approximately $0.05$ for renters, indicating growing monocentricity.
Importantly, at $\rho=1$ (the best match to the latest Census data) the spreading indices for renters and mortgagors already show some divergence. 

Thus, our results suggest that the polycentricity observed in our simulations is not a feature of the whole population. Instead, it arises from (i) the formation of a large centre, preferred by renters, in the proximity of the CBD, and (ii) the appearance of many smaller centres in the fringes, preferred by mortgagors.


\subsection{Geospatial distributions of the cost of housing}

\begin{figure}[t!]
    \centering
    \subfloat[]{
        \includegraphics[width=0.49\columnwidth,trim={1.1cm 1.3cm 0.3cm 1.1cm},clip]{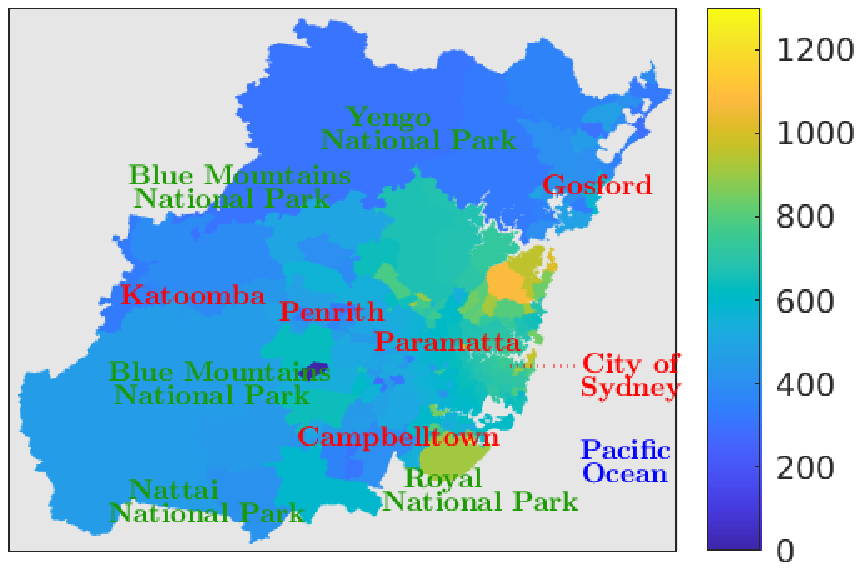}
        \label{fig:sydney-rent-2016}}
    \subfloat[]{
        \includegraphics[width=0.49\columnwidth,trim={1.1cm 0.9cm 0.3cm 0.6cm},clip]{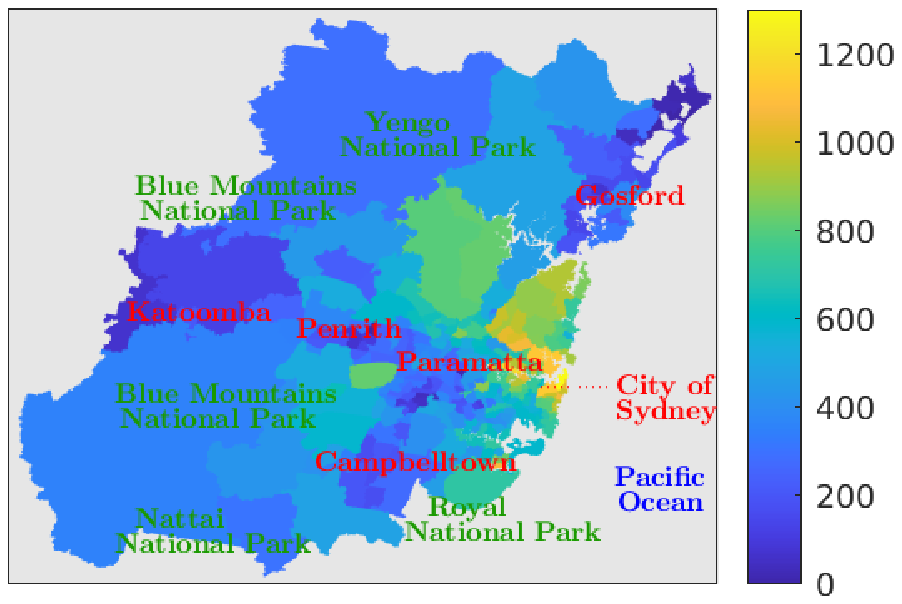}
        \label{fig:sydney-mort-2016}}\\
    \subfloat[]{
        \includegraphics[width=0.49\columnwidth,trim={1.1cm 0.8cm 0.3cm 0.6cm},clip]{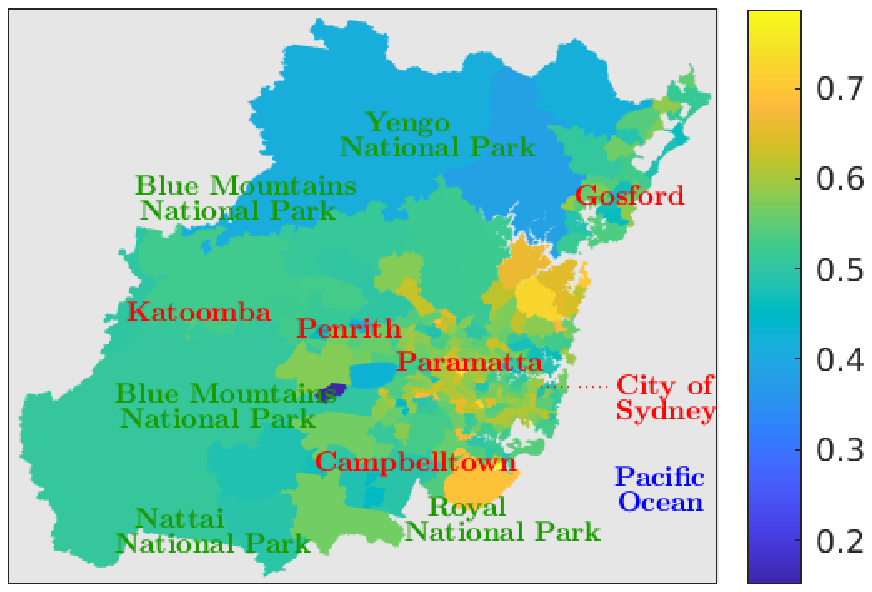}
        \label{fig:sydney-mort-rent-2016}}
    \subfloat[]{
        \includegraphics[width=0.49\columnwidth,trim={1.1cm 0.8cm 0.3cm 0.6cm},clip]{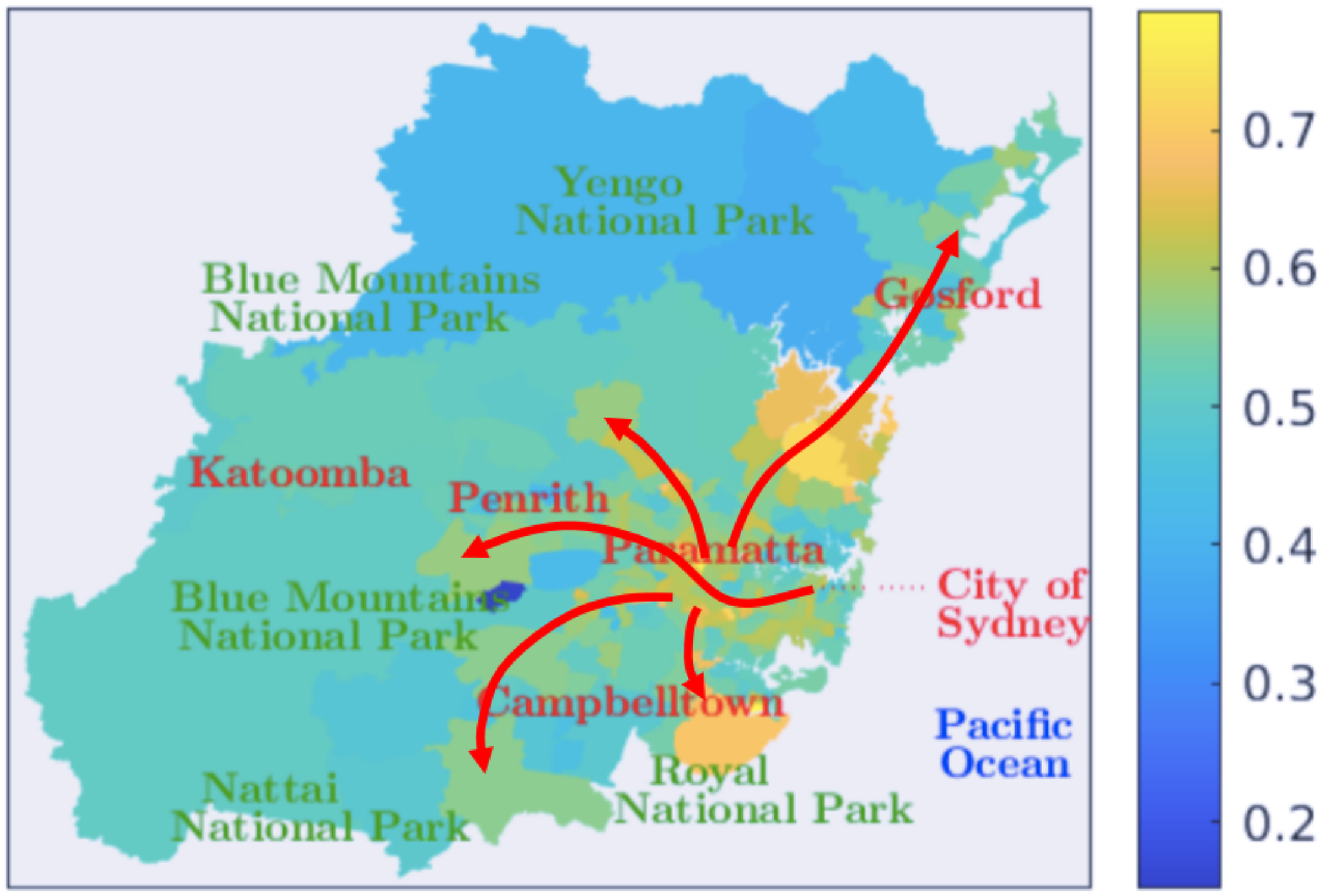} 
        \label{fig:sydney-mort-rent-2016-routes}}
    \caption{(a) Rental housing costs $R_j$ per week in Aus\$ as computed by the simulation ($\rho =  1$). (b) Mortgagor housing costs $M_j$ per week in Aus\$ as computed by the simulation ($\rho =  1$). (c) The ratio of rental costs to mortgage costs $R_j/M_j$. (d) The high value $R_j/M_j$ regions reflect the arterial routes through and out of the Greater Sydney region, an emergent phenomena of the model not apparent in individual plots of either $M_j$ or $R_j$.}
    \label{fig:sydney-mort-rent-2016-whole}
\end{figure}

Finally we want to understand the geographic distribution of the different costs of housing for renters and mortgagors, as opposed to the population distribution caused by housing attractiveness discussed in the previous section. Again, what we observe in Figure~\ref{fig:sydney-rent-2016} and~\ref{fig:sydney-mort-2016} is considerable heterogeneity in the distribution of the costs of housing, a tendency for renting costs to be higher when mortgage costs are lower, and vice versa, but these are not the same geographical distributions that we observe in Figure~\ref{fig:sydney-renters-mortgagors-2016}, indicating that population density, either of renters or mortgagors, is not a good indicator of where costs will be high or low, for either of these two groups.

Another phenomena emerges if we then take the ratio of rental costs to mortgage costs as in Figure~\ref{fig:sydney-mort-rent-2016} giving us a picture of the relative costs in the different regions throughout the Greater Sydney region. Here we can see some distinct patterns emerge from these distributions: along the arterial routes throughout Sydney, the rental costs have increased to nearly match those of the mortgagors’ costs, i.e. along these routes the cost of housing have nearly equilibriated for the renters and mortgagors. These arterial routes, shown in Figure~\ref{fig:sydney-mort-rent-2016-routes}, feed the northern, western, and southern regions of Sydney via the complex networks of road and rail services, see~\cite{GreaterSydneyCommission2016} and Appendix~\ref{GSTrafficFlows}. The implication is that, without feeding the model anything related to the short-term geospatial patterns of human movement or long-term population settlement, the cost of housing forms equilibrium patterns along transport routes and this is one of the notable emergent phenomenon of this model. These costs are also approximately within the expected numerical range we would observe due to the housing sales and rental data across Sydney. 


\section{Discussion}

A city is a large, complex dynamical system, the evolution of which is important as policy makers seek to satisfy multiple constraints across many different interest groups, often with conflicting expectations. Two of the most important interest groups are the residents (workers) and the industries that employ them, both of which are supported by infrastructure such as roads, but this infrastructure needs to be balanced against the quality of life aspects residents need in order to want to live in a city.

In this article we have shown that by starting with an optimisation over existing commuter networks and job locations (using MaxEnt) we are able to identify a plausible evolution of the population's demand for resettling in a modern, industrialised, city. In this approach, the known population centres emerge naturally, while the cost of housing estimates for both renters and mortgage holders in dollar value develop in line with the data. In addition, the housing patterns are found to  mirror major road networks. These findings provide an illustrative example of the usefulness of modern complex systems tools as inputs for policy making at the macro-economic scale of a city using currently available data-sets.

The result is a model that uses high resolution socio-economic data to represent the competing forces that drive urban settlement patterns: the costs of housing (for renters and mortgage holders), changes in preferences based on population density, location, earning potential, and the transport costs to and from work. The cost of transport can be thought of as a type of friction associated with the flow of income from the location where the income is earned to the mortgage or rental payments (residential location).

Our results show that non-trivial spatial structures emerge through these competitive processes. We see that, if the housing markets were to relax from their current monocentric state~\cite{slavko2019dynamic} to an equilibrium state, the populations of both Sydney and Melbourne would, in the long run, settle at the polycentric side of a sharp transition caused by factors influencing the attractiveness of regions within a city~\cite{muneepeerakul2012critical}.
A similar transition was found in a study of the income flow (from suburbs of employment to suburbs of residence) within Greater Sydney~\cite{crosato2018critical}.
In that study, the suburbs' attractiveness was defined in terms of the availability of various services (e.g., measured by proximity to supermarkets and grocery stores), rather than the rent and mortgage expenses.
It was reported that spatial configurations of the income flow would relax towards a monocentric configuration, although such configuration was close, in the parameter space, to the critical regime transitioning to polycentricity.

Importantly, our results reveal significant differences between renters and mortgagors; the renters are more attracted to central business districts, while mortgagors are repelled. A curious consequence of this dynamic is that the ratio of mortgage to rental costs in each suburb appear to be tending, in the long run of these simulations, towards a fairly even ratio that follows the arterial traffic routes through the city. This appears to be an entirely new outcome from a model of this type and warrants closer investigation in future research.

\section*{Acknowledgement}
Sydney Informatics Hub at the University of Sydney provided access to HPC computational resources
that have contributed to the research results reported within the paper.



\appendix

\section{Attractiveness functions}
\label{sec:attractiveness-functions}

A set of different attractiveness functions have been tested and compared based on the error

\begin{equation}
e^\text{*} = 
\frac{
\sum^{N^\text{emp}}_{i=1}\sum^{N^\text{res}}_{j=1}
\frac{1}{2}\left|{\cal T}^\text{*(start)}_{ij} - T_{ij}^\text{*}\right|
}
{\sum^{N^\text{emp}}_{i=1}\sum^{N^\text{res}}_{j=1} {\cal T}_{ij}^\text{*}} ,
\end{equation}
between the commuting matrix given by the Census ${\cal T}^\text{*}_{ij}$ and the one predicted by the model at the first iteration $T^\text{*(start)}_{ij}$, where `*' is either `R', indicating renters, or `M', indicating mortgagors.
The results are summarised Table~\ref{tab:attract} for the case of Greater Sydney in year 2016.

\begin{figure}[h!]
    \centering
    \includegraphics[width=0.75\textwidth]{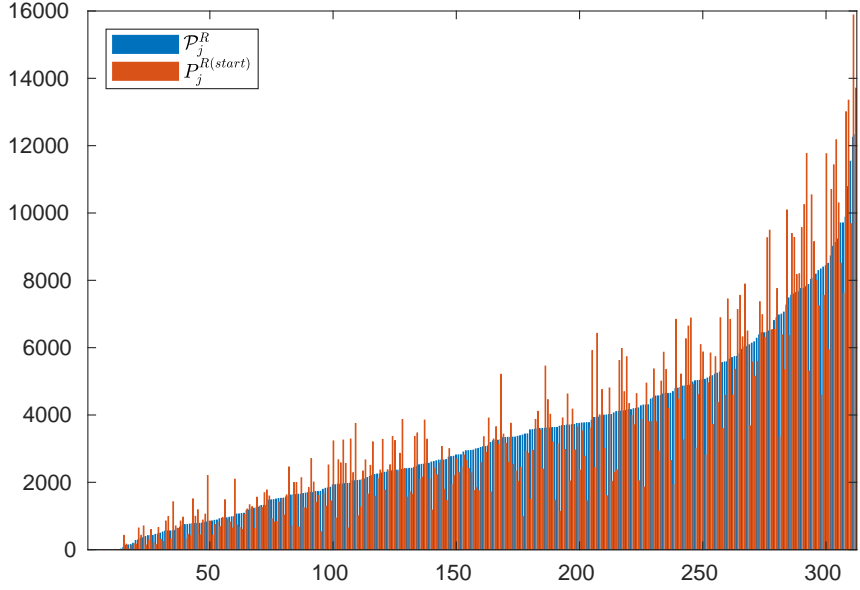}
    \caption{Number of renters per suburb: Census data vs MaxEnt. Each blue bar represents the number of renters in a suburb ${\cal P}_j^\text{R}$ as per Census data.
    The suburbs are sorted from the least populated (in terms of renters) to the most populated.
    The red bars represent the number of renters $P_j^\text{R(start)}$ predicted by MaxEnt using $A_j=\log(1+(\hat{R}-R_j)(\hat{M}-M_j)P_j)$, in the same order as the blue bars.}
    \label{fig:hist-rent}
\end{figure}

\begin{figure}[h!]
    \centering
    \includegraphics[width=0.75\textwidth]{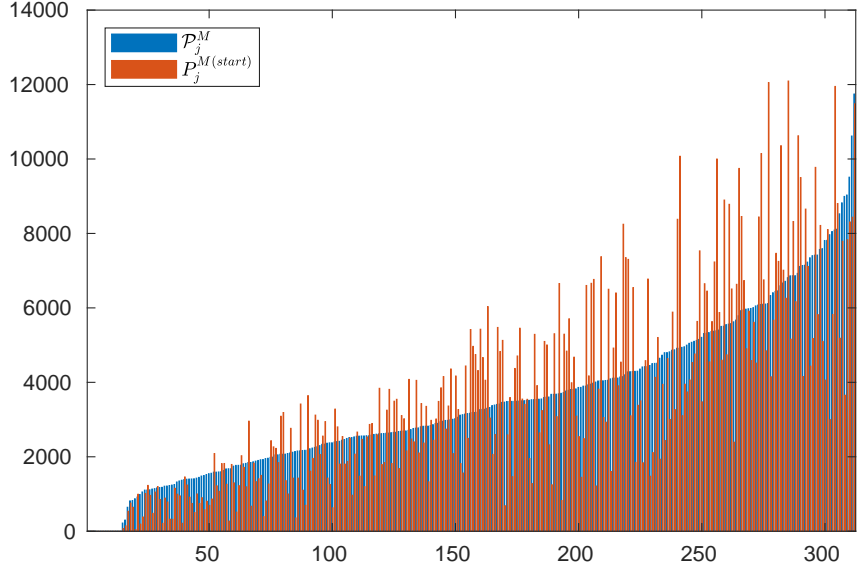}
    \caption{Number of mortgagors per suburb: Census data vs MaxEnt. Each blue bar represents the number of mortgagors in a suburb ${\cal P}_j^\text{M}$ as per Census data.
    The suburbs are sorted from the least populated (in terms of mortgagors) to the most populated.
    The red bars represent the number of mortgagors $P_j^\text{M(start)}$ predicted by MaxEnt using $A_j=\log(1+(\hat{R}-R_j)(\hat{M}-M_j)P_j^\text{M})$, in the same order as the blue bars.}
    \label{fig:hist-mort}
\end{figure}

The best matching attractiveness is $A_j=\log(1+(\hat{R}-R_j)(\hat{M}-M_j)P_j)$ for renters and $A_j=\log(1+(\hat{R}-R_j)(\hat{M}-M_j)P^\text{M}_j)$ for mortgagors.
Fig.~\ref{fig:hist-rent} and \ref{fig:hist-mort} compare the population of each suburb according to the Census data with the one predicted by MaxEnt using these attractiveness functions.

\begin{table}[h!]
\centering
\begin{tabular}{|c|c|c|c|c|c|c|}
\hline
Attractiveness & $e^\text{R}$ & $e^\text{M}$ & $\hat{\alpha}^\text{R}$ & $\hat{\alpha}^\text{M}$ & $\hat{\gamma}^\text{R}$ & $\hat{\gamma}^\text{M}$ \\
\hline
$A_j=\log(1+\text{rand})$ & 0.3379 & 0.4206 & 0.0 & 0.0 & 0.024 & 0.023\\
$A_j=\log(1+R_j)$ & 0.3311 & 0.4203 & -0.8 & 0.2 & 0.025 & 0.024\\
$A_j=\log(1+M_j)$ & 0.3264 & 0.4206 & -1.4 & 0.0 & 0.025 & 0.023\\
$A_j=\log(1+\hat{R}-R_j)$ & 0.3311 & 0.4205 & 2.1 & -0.3 & 0.025 & 0.023\\
$A_j=\log(1+\hat{M}-M_j)$ & 0.3266 & 0.4204 & 2.8 & 0.4 & 0.026 & 0.023\\
$A_j=\log(1+R_jP^\text{R}_j)$ & 0.3069 & 0.4163 & 0.4 & 0.1 & 0.023 & 0.023\\
$A_j=\log(1+R_jP_j)$ & 0.3130 & 0.4007 & 0.5 & 0.5 & 0.022 & 0.023\\
$A_j=\log(1+M_jP^\text{M}_j)$ & 0.3225 & 0.3592 & 0.2 & 0.9 & 0.023 & 0.022\\
$A_j=\log(1+M_jP_j)$ & 0.3122 & 0.4016 & 0.5 & 0.5 & 0.022 & 0.023\\
$A_j=\log(1+(\hat{R}-R_j)P^\text{R}_j)$ & 0.2918 & 0.4164 & 0.6 & 0.1 & 0.022 & 0.023\\
$A_j=\log(1+(\hat{R}-R_j)P_j)$ & 0.2872 & 0.3994 & 1.0 & 0.6 & 0.022 & 0.023\\
$A_j=\log(1+(\hat{M}-M_j)P^\text{M}_j)$ & 0.3112 & 0.3297 & 0.5 & 1.4 & 0.024 & 0.020\\
$A_j=\log(1+(\hat{M}-M_j)P_j)$ & 0.2857 & 0.3982 & 1.0 & 0.7 & 0.022 & 0.022\\
$A_j=\log(1+R_jM_j)$ & 0.3282 & 0.4250 & -0.6 & 0.1 & 0.025 & 0.024\\
$A_j=\log(1+R_jM_jP_j)$ & 0.3222 & 0.4039 & 0.3 & 0.4 & 0.023 & 0.024\\
$A_j=\log(1+(\hat{R}-R_j)(\hat{M}-M_j))$ & 0.3279 & 0.4206 & 1.4 & 0.0 & 0.025 & 0.023\\
$A_j=\log(1+(\hat{R}-R_j)(\hat{M}-M_j)P_j)$ & \textbf{0.2807} & 0.3991 & \textbf{1.0} & 0.6 & \textbf{0.022} & 0.022\\
$A_j=\log(1+(\hat{R}-R_j)(\hat{M}-M_j)P^\text{R}_j)$ & 0.2885 & 0.4152 & 0.7 & 0.1 & 0.022 & 0.023\\
$A_j=\log(1+(\hat{R}-R_j)(\hat{M}-M_j)P^\text{M}_j)$ & 0.3076 & \textbf{0.3282} & 0.5 & \textbf{1.4} & 0.024 & \textbf{0.020}\\
$A_j=\log(1+S_j)$ & 0.3210 & 0.4173 & 0.2 & 0.1 & 0.023 & 0.023\\
$A_j=\log(1+S_jP_j)$ & 0.3062 & 0.4102 & 0.2 & 0.1 & 0.022 & 0.023\\
$A_j=\log(1+S_jP^\text{R}_j)$ & 0.3035 & 0.4171 & 0.2 & 0.1 & 0.022 & 0.023\\
$A_j=\log(1+S_jP^\text{M}_j)$ & 0.3089 & 0.3952 & 0.2 & 0.3 & 0.023 & 0.021\\
\hline
\end{tabular}
\caption{Summary of the considered attractiveness functions, including the the errors on the mortgagors and renters population and the relative Lagrangians values.}
\label{tab:attract}
\end{table}

\newpage
\ 
\newpage
\section{Calibration of the Lagrangians}
\label{sec:calibration}

\begin{figure}[h!]
    \centering
    \includegraphics[width=0.65\columnwidth, trim={0.0cm 0.1cm 0.8cm 0.5cm},clip]{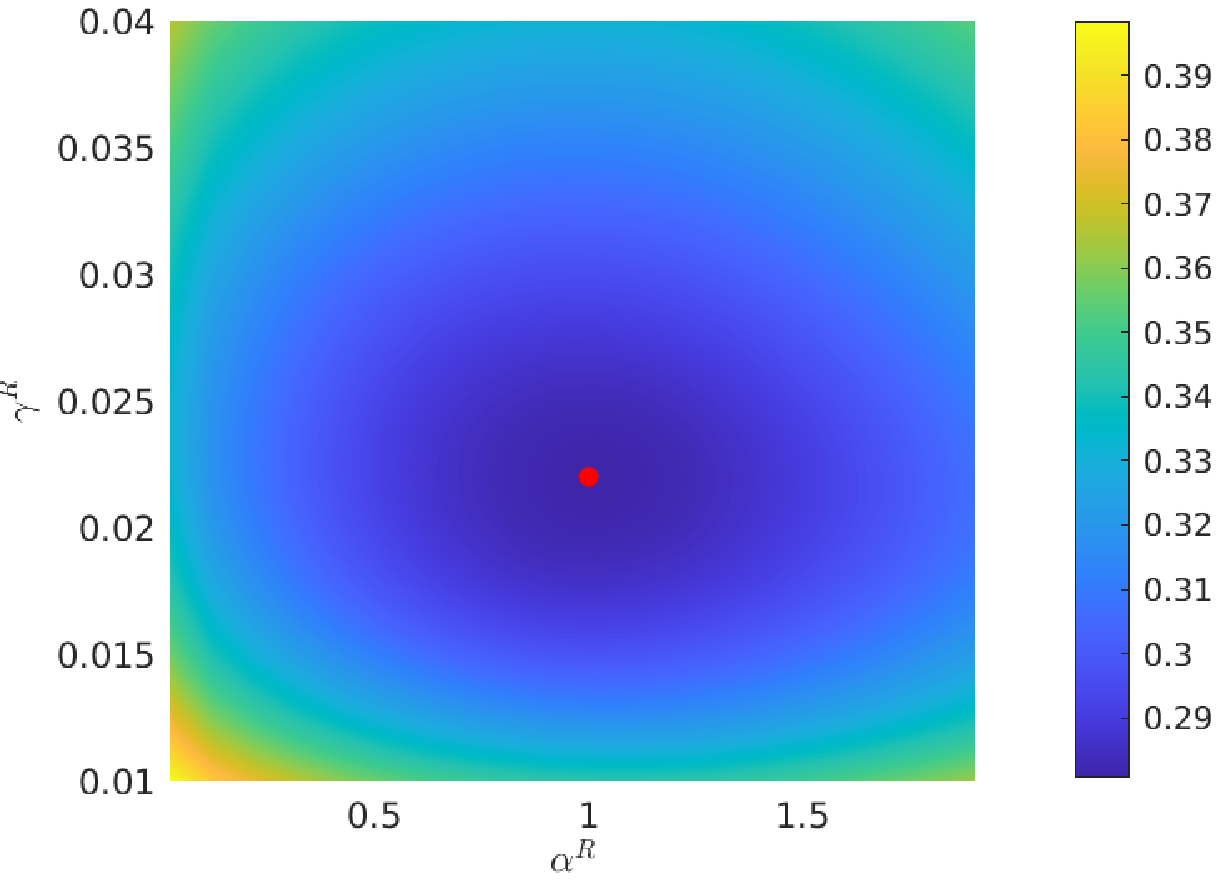}\\
\includegraphics[width=0.65\columnwidth, trim={0.0cm 0.1cm 0.8cm 0.1cm},clip]{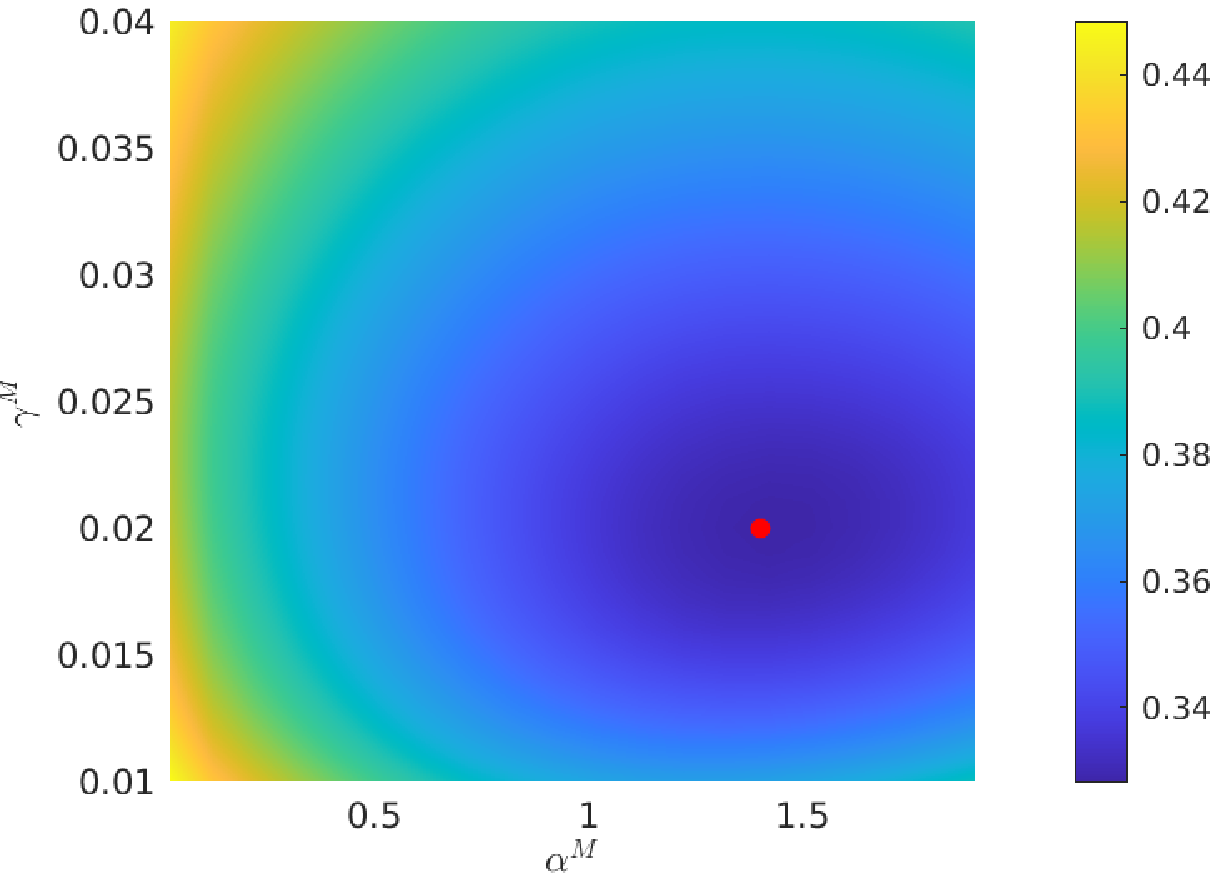}
    \caption{Calibration of the Lagrangians for Sydney 2016.
    The horizontal axis represents $\alpha^\text{R}$ (top) or $\alpha^\text{M}$ (bottom), while the vertical axis represents $\gamma^\text{R}$ (left) or $\gamma^\text{M}$ (right).
    The color represents the difference between the travel-to-work matrices yielded by MaxEnt and the travel-to-work matrices obtained from the census.
    The red dots indicates the combination $\{\hat{\alpha}^\text{*}, \hat{\gamma}^\text{*}\}$ that best matches the Census data.}
    \label{fig:calibration}
\end{figure}

The calibration process consisted in the numerical estimation of the optimal values of $\alpha^\text{R}$, $\gamma^\text{R}$, $\alpha^\text{M}$ and $\gamma^\text{M}$, i.e., the values that minimise the difference between the travel-to-work matrices $T_{ij}^\text{R(start)}$ and $T_{ij}^\text{M(start)}$ yielded by MaxEnt and the actual travel-to-work matrices ${\cal T}_{ij}^\text{R}$ and ${\cal T}_{ij}^\text{M}$:
\begin{equation}
    \{\hat{\alpha}^\text{*}, \hat{\gamma}^\text{*}\} = \arg\min_{\alpha^\text{*},\gamma^\text{*}} \sum_i \sum_j | T_{ij}^\text{*(start)} - {\cal T}_{ij}^\text{*}| ,
\end{equation}
where `*' is either `R', indicating renters, or `M', indicating mortgagors.
The result is summarised in Fig.~\ref{fig:calibration}.

\newpage
\section{Greater Sydney 2011}

\begin{figure*}[h!]
    \centering
    \subfloat[]{
        \includegraphics[width=0.45\columnwidth,trim={1.1cm 0.9cm 0.8cm 0.3cm},clip]{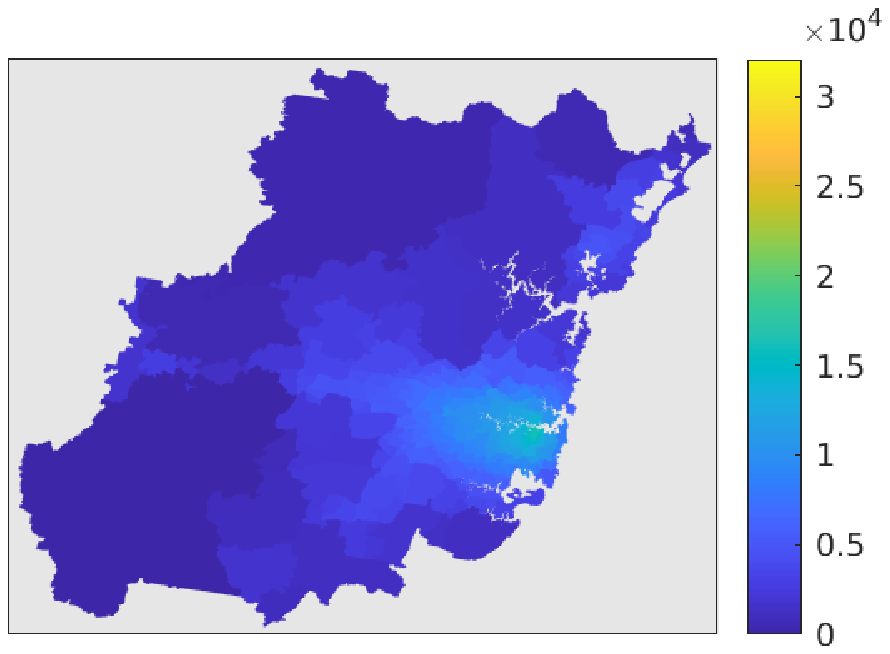}
        \label{fig:sydney-maps-2011-sprawl}}
    \subfloat[]{
        \includegraphics[width=0.45\columnwidth,trim={1.1cm 0.9cm 0.8cm 0.3cm},clip]{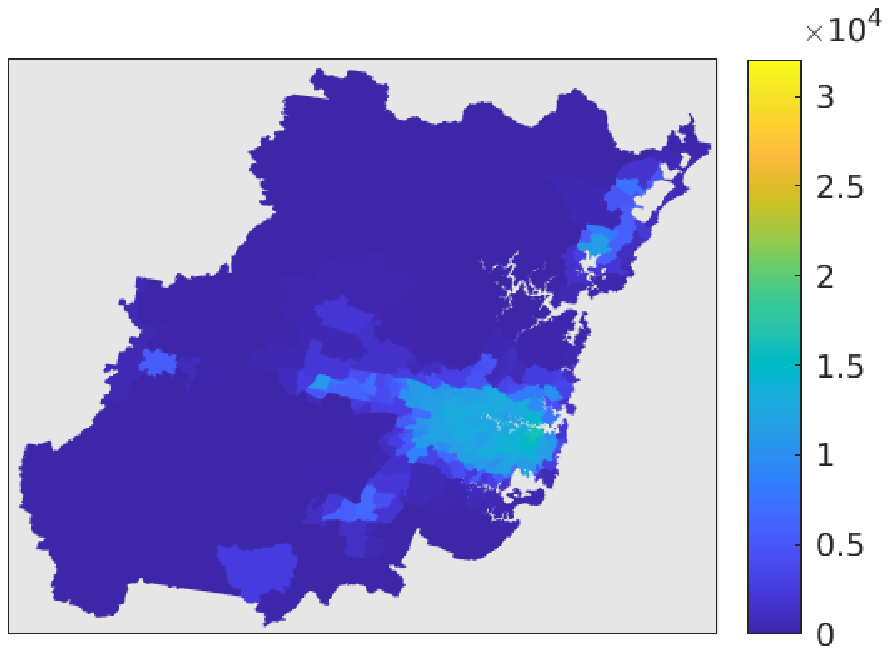}
        \label{fig:sydney-maps-2011-critical}}\\
    \subfloat[]{
        \includegraphics[width=0.45\columnwidth,trim={1.1cm 0.9cm 0.8cm 0.3cm},clip]{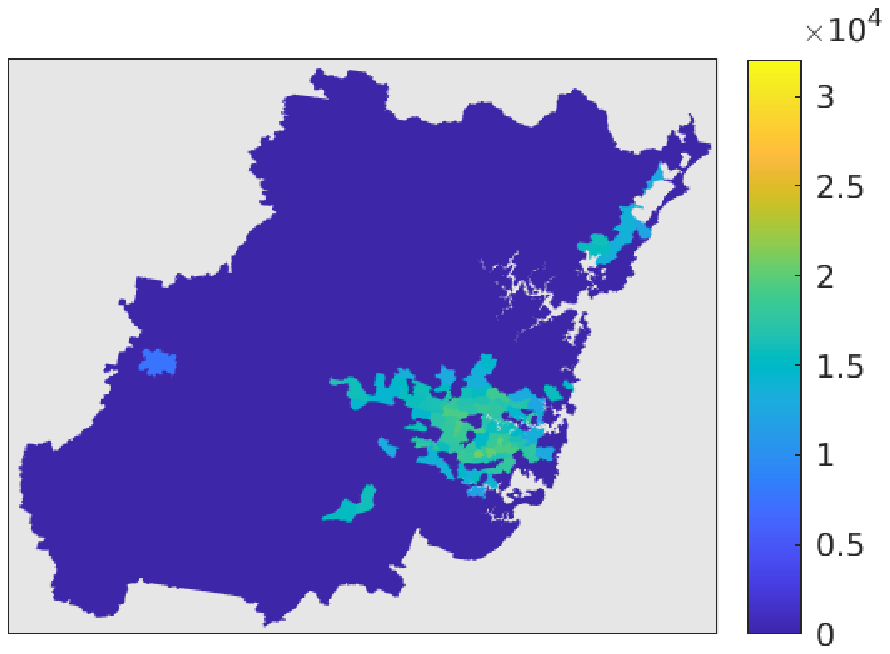}
        \label{fig:sydney-maps-2011-poly}}
    \subfloat[]{
        \includegraphics[width=0.45\columnwidth,trim={1.1cm 0.9cm 0.7cm 0.3cm},clip]{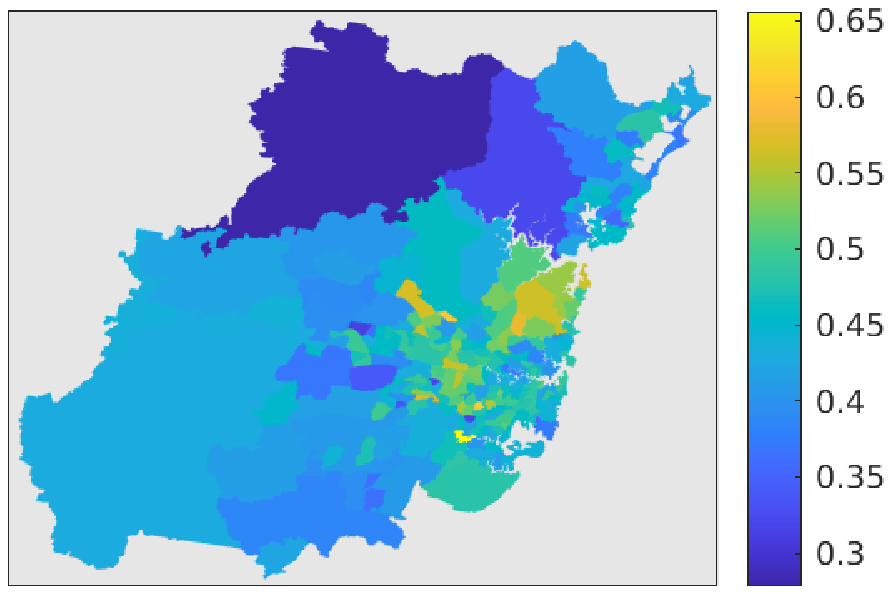}
        \label{fig:sydney-maps-2011-rent-to-mort}}\\
    \subfloat[]{
        \includegraphics[width=0.45\columnwidth,trim={1.1cm 0.9cm 0.7cm 0.3cm},clip]{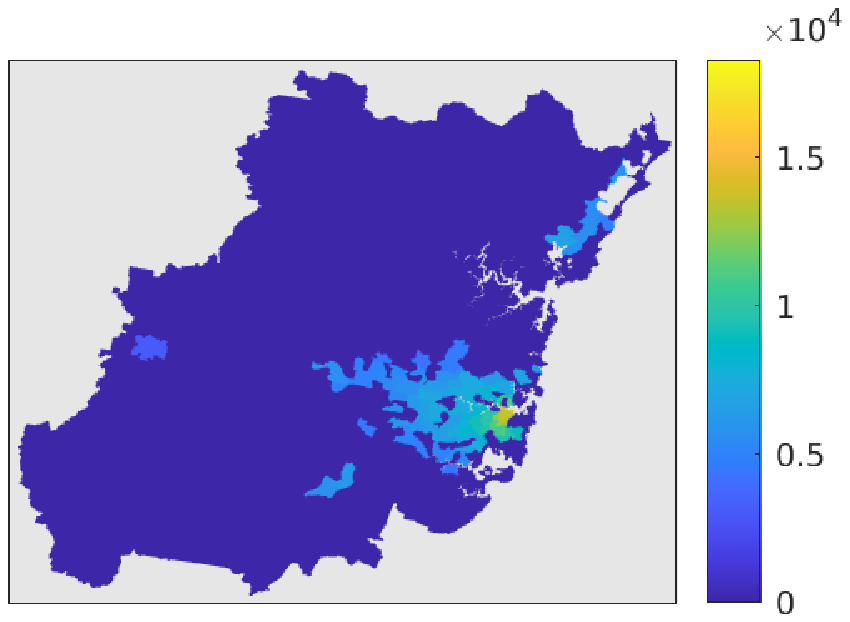}
        \label{fig:sydney-maps-2011-rent}}
    \subfloat[]{
        \includegraphics[width=0.45\columnwidth,trim={1.1cm 0.9cm 0.7cm 0.3cm},clip]{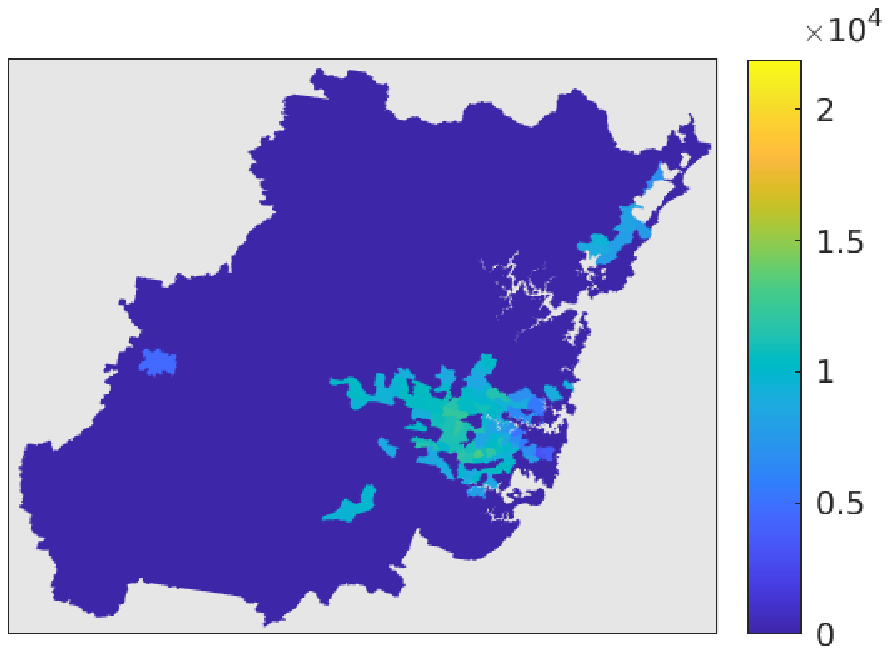}
        \label{fig:sydney-maps-2011-mort}}%
    \caption{Greater Sydney 2011. (a-c) Distribution of the population $P_j$ over the residential suburbs $j$ after running the BLV dynamics using (a) $\rho=0.2$, (b) $\rho=0.6$ and (c) $\rho=1$
    (d) Ratio of the rental costs to the mortgage costs after running the BLV dynamics with $\rho=1$.
    (e-f) Distribution of the renters population $P^\text{R}_j$ (e) and the mortgagors population $P^\text{M}_j$ (f) over the residential suburbs $j$ after running the BLV dynamics with $\rho=1$.}
    \label{fig:sydney-maps-2011}
\end{figure*}


\newpage
\section{Greater Melbourne 2016}

\begin{figure*}[h!]
    \centering
    \subfloat[]{
        \includegraphics[width=0.42\columnwidth,trim={1.1cm 0.7cm 0.8cm 0.1cm},clip]{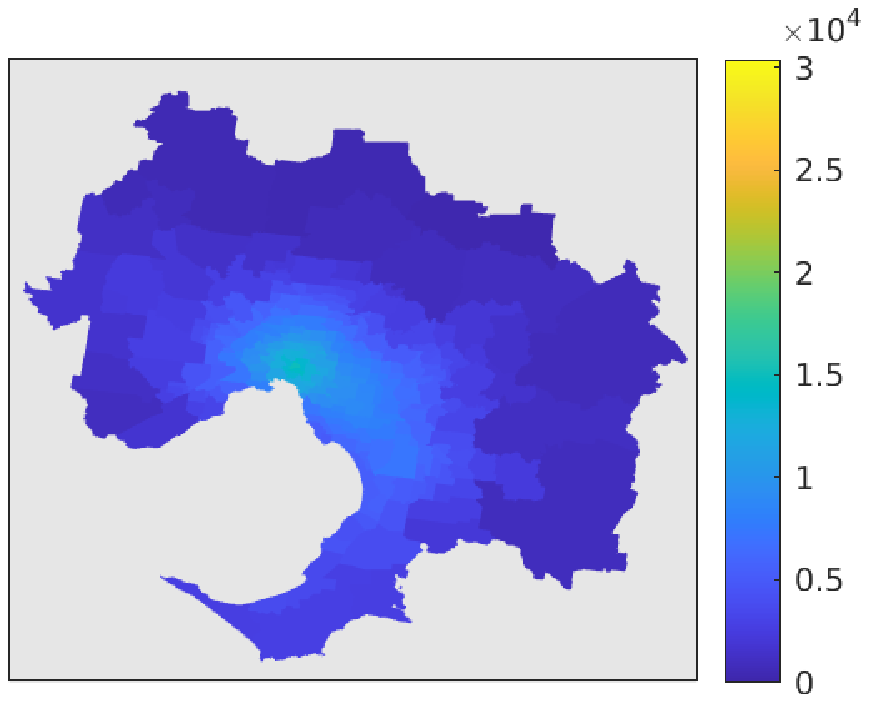}
        \label{fig:melbourne-maps-2016-sprawl}}
    \subfloat[]{
        \includegraphics[width=0.42\columnwidth,trim={1.1cm 0.7cm 0.8cm 0.1cm},clip]{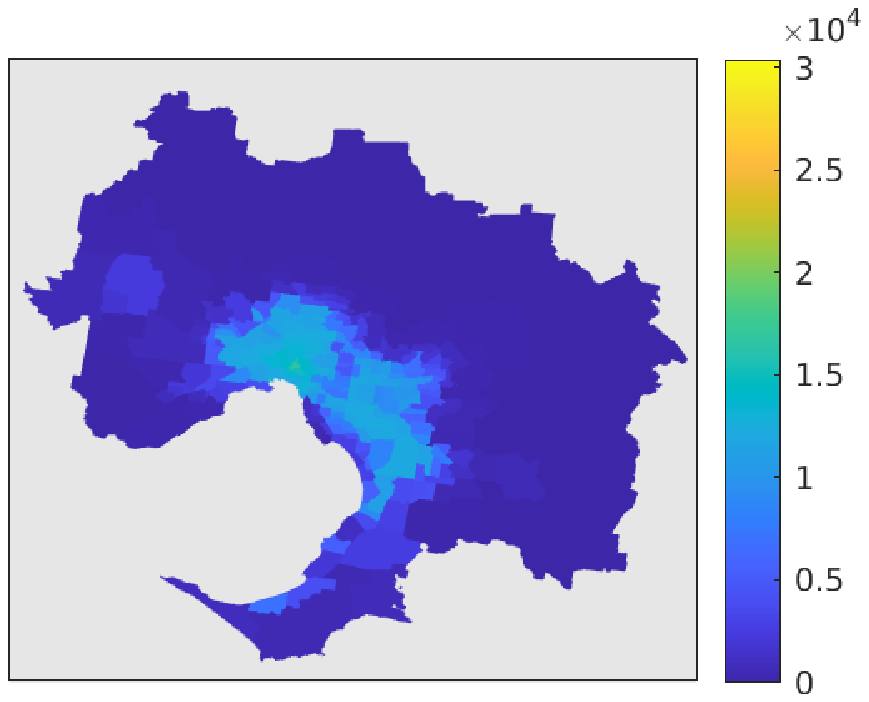}
        \label{fig:melbourne-maps-2016-critical}}\\
    \subfloat[]{
        \includegraphics[width=0.42\columnwidth,trim={1.1cm 0.7cm 0.8cm 0.1cm},clip]{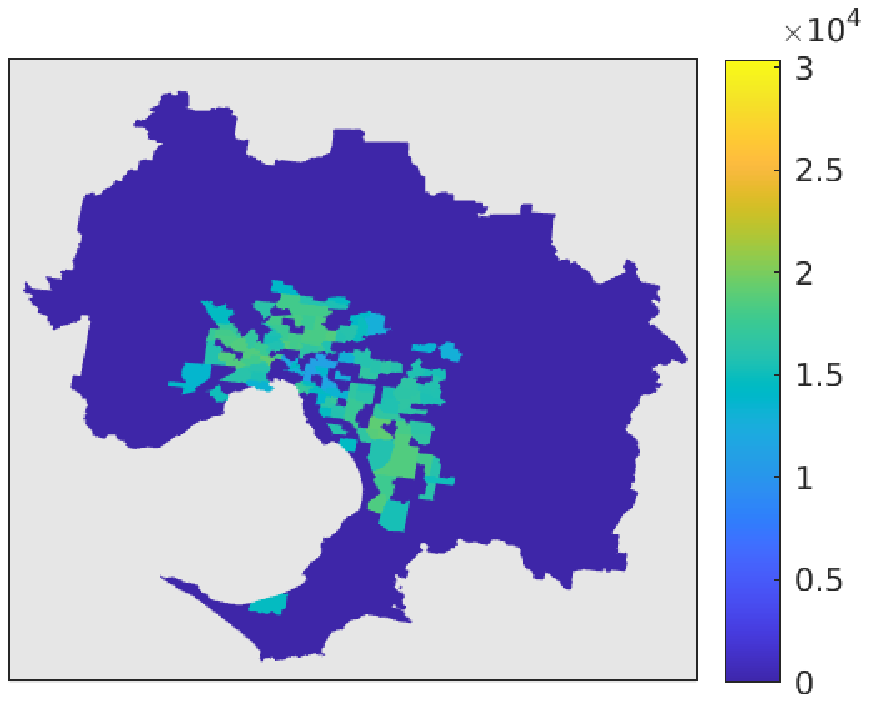}
        \label{fig:melbourne-maps-2016-poly}}
    \subfloat[]{
        \includegraphics[width=0.42\columnwidth,trim={1.1cm 0.7cm 0.7cm 0.3cm},clip]{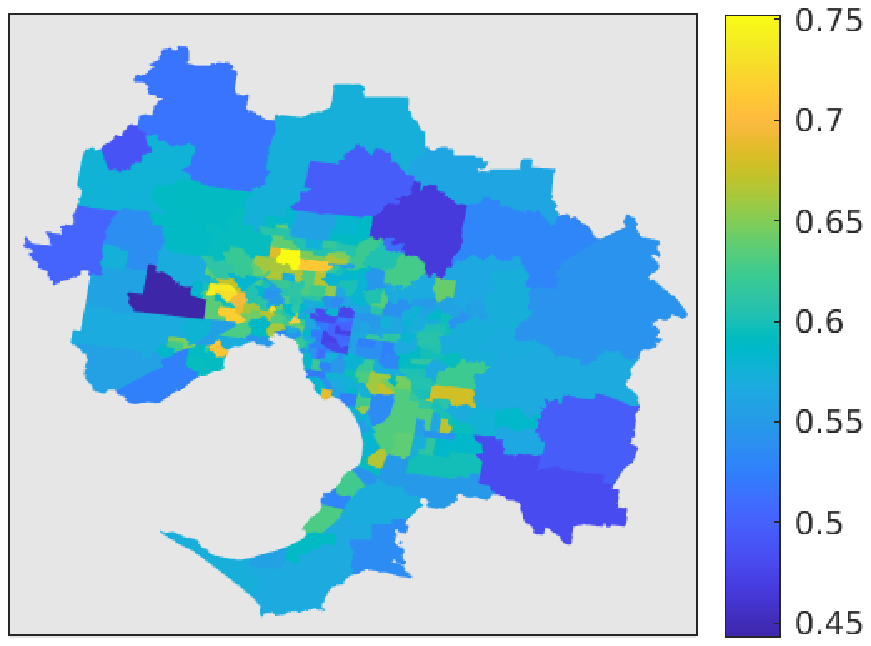}
        \label{fig:melbourne-maps-2016-rent-to-mort}}\\
    \subfloat[]{
        \includegraphics[width=0.42\columnwidth,trim={1.1cm 0.7cm 0.7cm 0.0cm},clip]{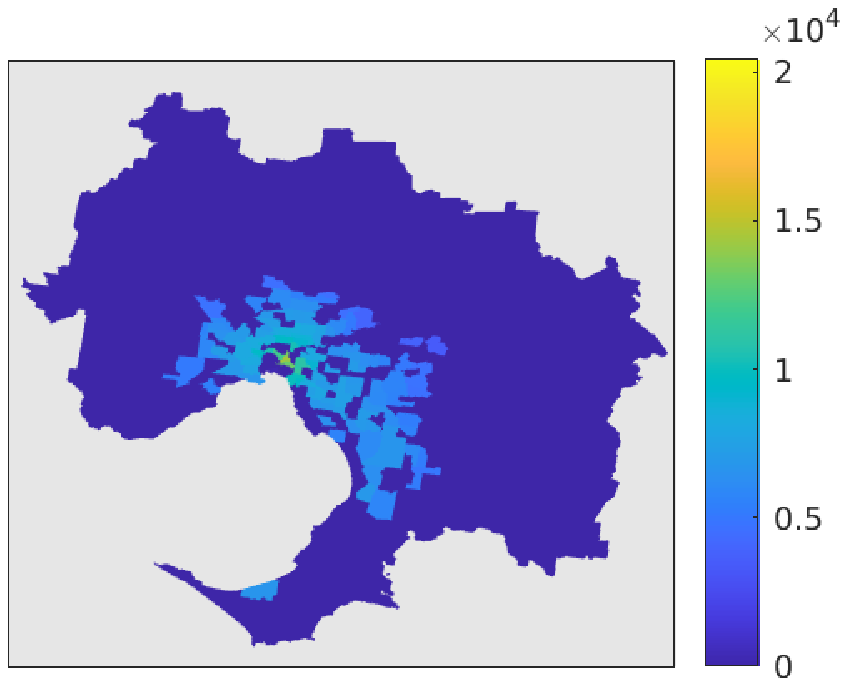}
        \label{fig:melbourne-maps-2016-rent}}
    \subfloat[]{
        \includegraphics[width=0.42\columnwidth,trim={1.1cm 0.7cm 0.7cm 0.0cm},clip]{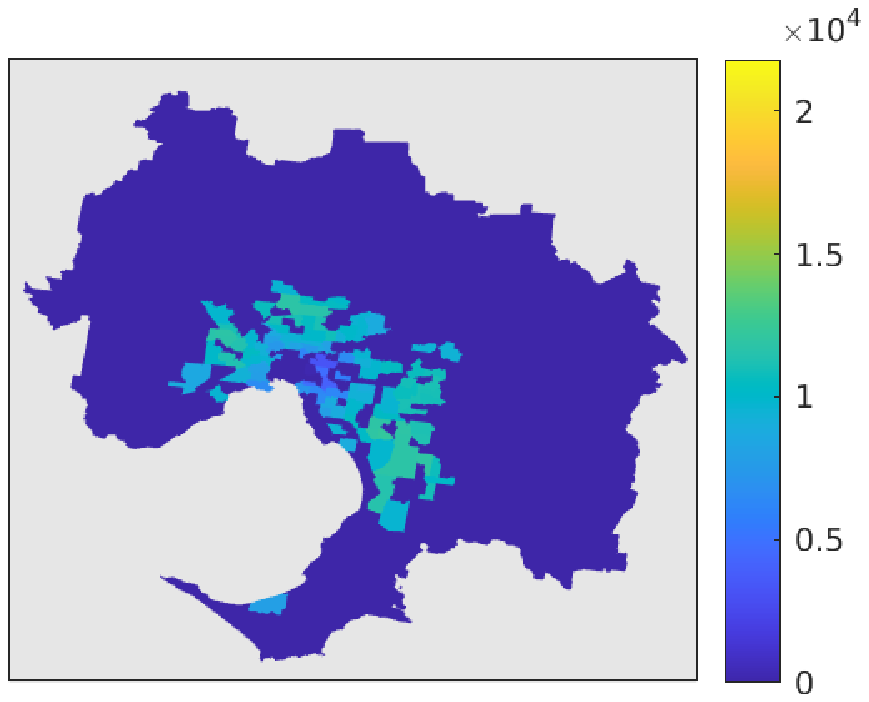}
        \label{fig:melbourne-maps-2016-mort}}%
    \caption{Greater Melbourne 2016. (a-c) Distribution of the population $P_j$ over the residential suburbs $j$ after running the BLV dynamics using (a) $\rho=0.2$, (b) $\rho=0.6$ and (c) $\rho=1$
    (d) Ratio of the rental costs to the mortgage costs after running the BLV dynamics with $\rho=1$.
    (e-f) Distribution of the renters population $P^\text{R}_j$ (e) and the mortgagors population $P^\text{M}_j$ (f) over the residential suburbs $j$ after running the BLV dynamics with $\rho=1$.}
    \label{fig:melbourne-maps-2016}
\end{figure*}


\newpage
\section{Greater Melbourne 2011}

\begin{figure*}[h!]
    \centering
    \subfloat[]{
        \includegraphics[width=0.42\columnwidth,trim={1.1cm 0.7cm 0.8cm 0.1cm},clip]{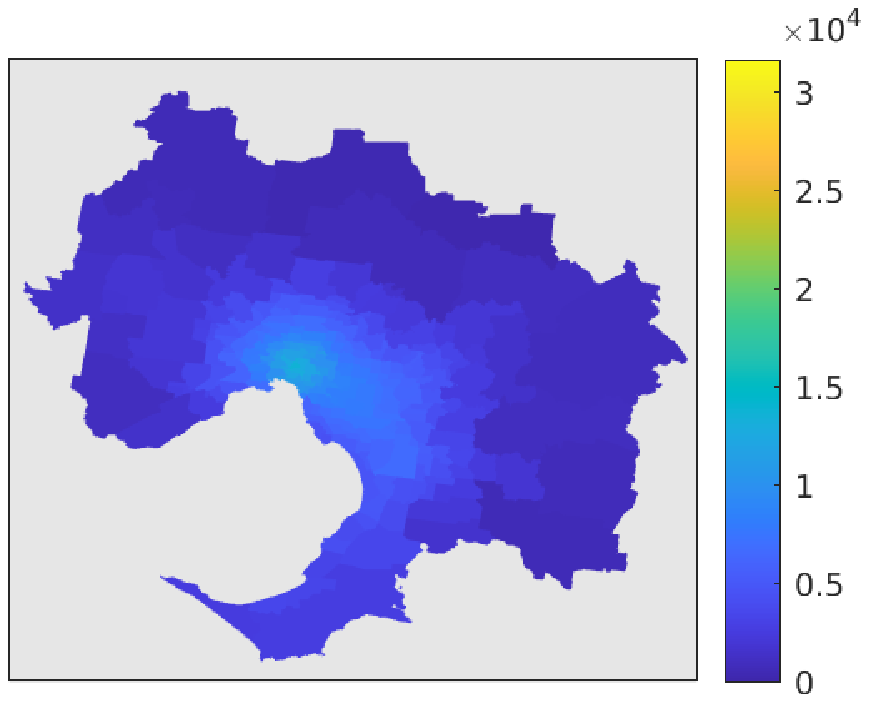}
        \label{fig:melbourne-maps-2011-sprawl}}
    \subfloat[]{
        \includegraphics[width=0.42\columnwidth,trim={1.1cm 0.7cm 0.8cm 0.1cm},clip]{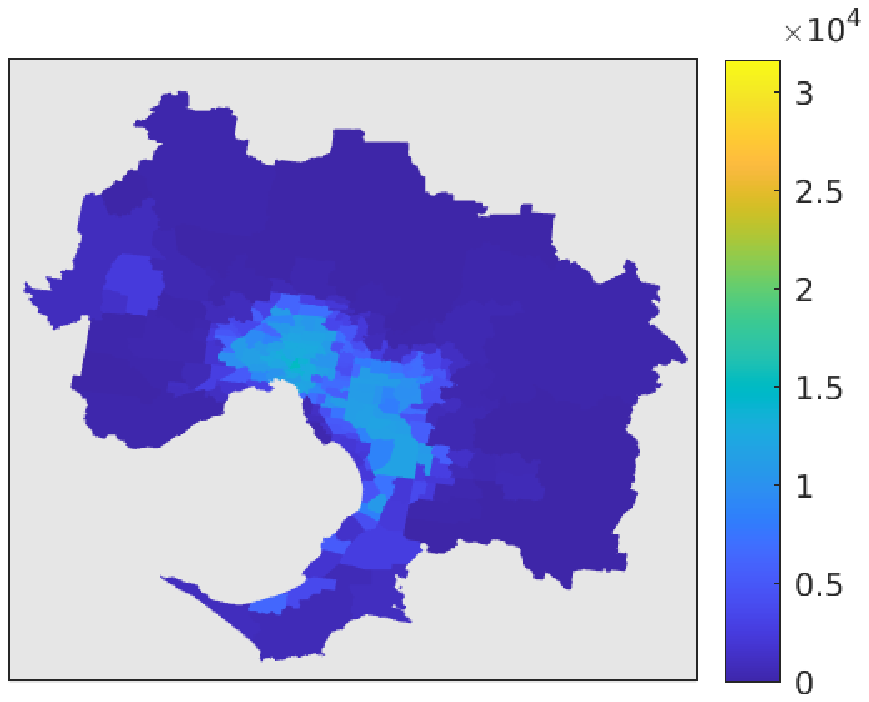}
        \label{fig:melbourne-maps-2011-critical}}\\
    \subfloat[]{
        \includegraphics[width=0.42\columnwidth,trim={1.1cm 0.7cm 0.8cm 0.1cm},clip]{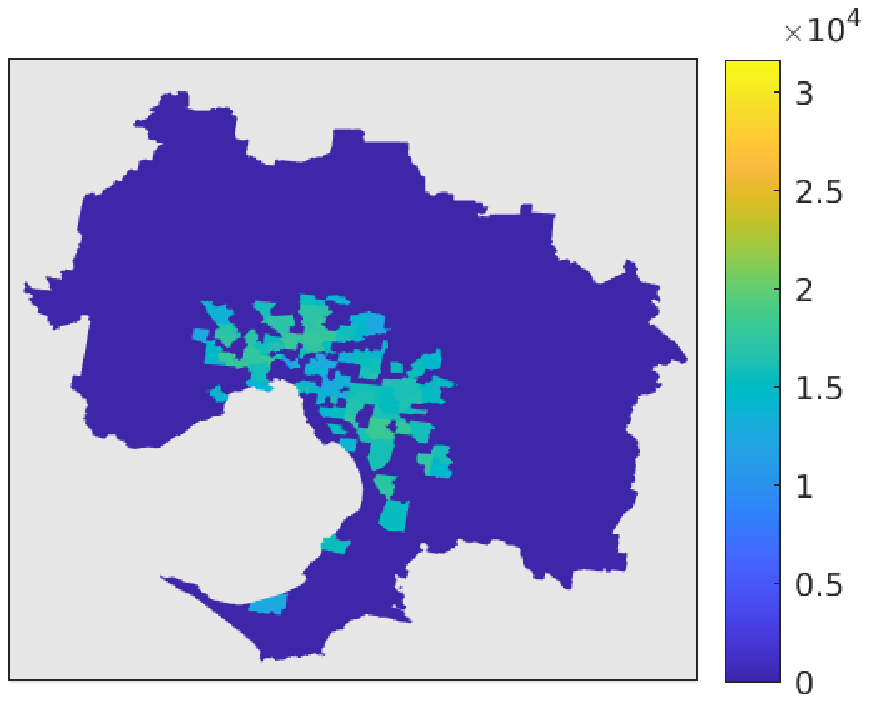}
        \label{fig:melbourne-maps-2011-poly}}
    \subfloat[]{
        \includegraphics[width=0.42\columnwidth,trim={1.1cm 0.7cm 0.7cm 0.3cm},clip]{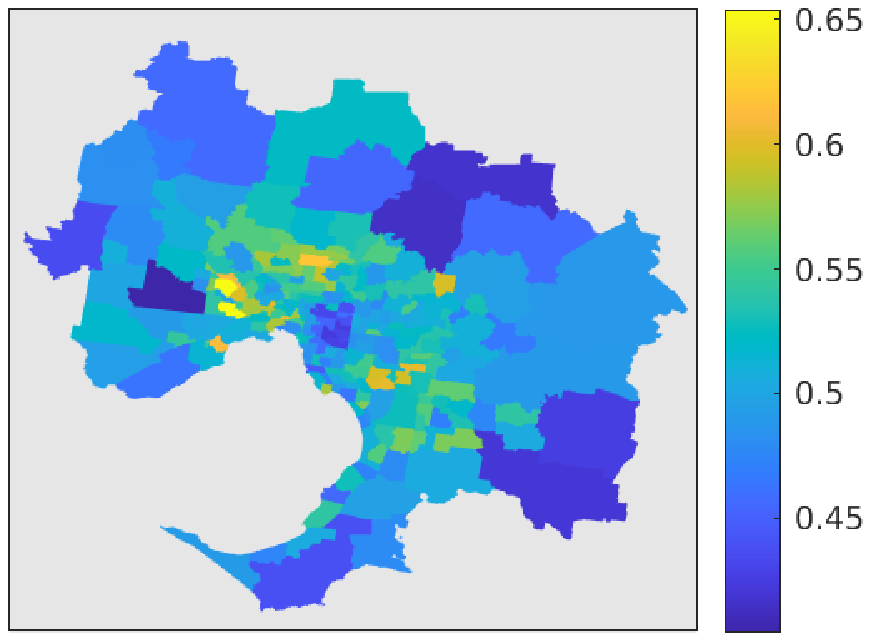}
        \label{fig:melbourne-maps-2011-rent-to-mort}}\\
    \subfloat[]{
        \includegraphics[width=0.42\columnwidth,trim={1.1cm 0.7cm 0.7cm 0.0cm},clip]{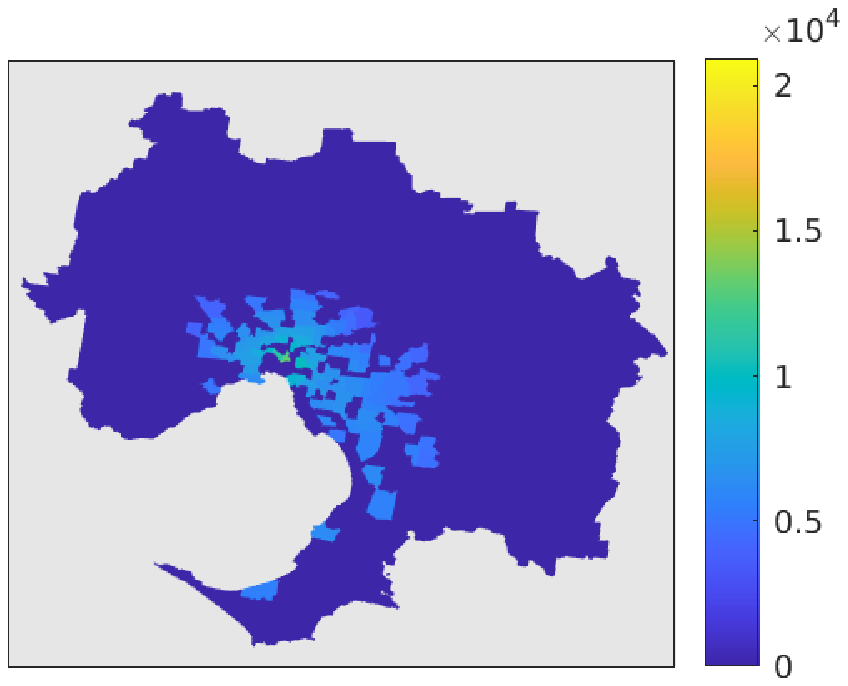}
        \label{fig:melbourne-maps-2011-rent}}
    \subfloat[]{
        \includegraphics[width=0.42\columnwidth,trim={1.1cm 0.7cm 0.7cm 0.0cm},clip]{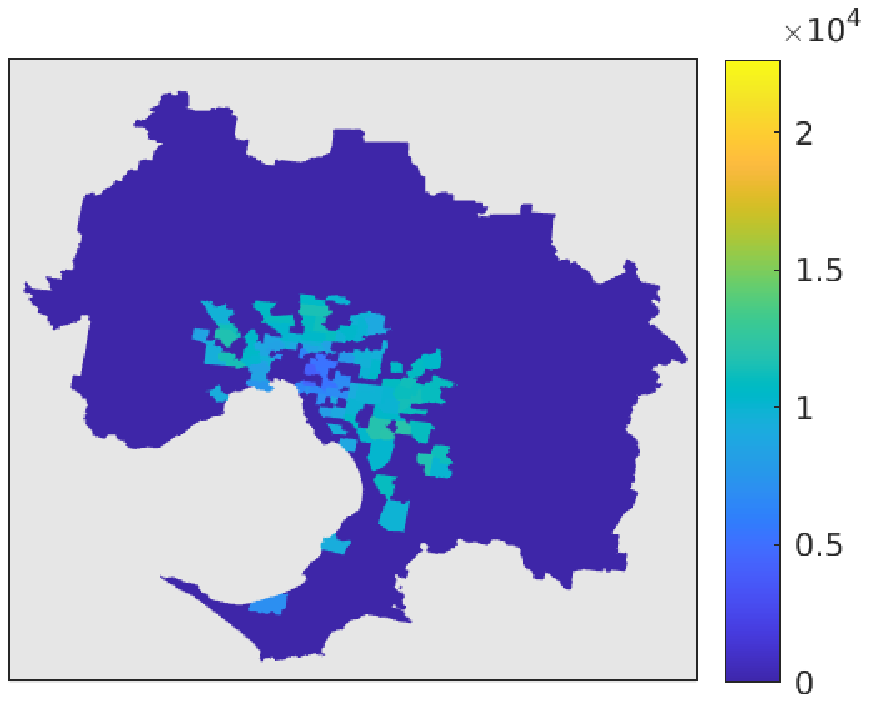}
        \label{fig:melbourne-maps-2011-mort}}%
    \caption{Greater Melbourne 2011. (a-c) Distribution of the population $P_j$ over the residential suburbs $j$ after running the BLV dynamics using (a) $\rho=0.2$, (b) $\rho=0.6$ and (c) $\rho=1$
    (d) Ratio of the rental costs to the mortgage costs after running the BLV dynamics with $\rho=1$.
    (e-f) Distribution of the renters population $P^\text{R}_j$ (e) and the mortgagors population $P^\text{M}_j$ (f) over the residential suburbs $j$ after running the BLV dynamics with $\rho=1$.}
    \label{fig:melbourne-maps-2011}
\end{figure*}

\newpage
\section{Greater Sydney Traffic Flows \label{GSTrafficFlows}}

\begin{figure*}[h!]
    \centering
    \includegraphics[width=0.75\columnwidth,trim={1.1cm 0.9cm 0.8cm 0.1cm},clip]{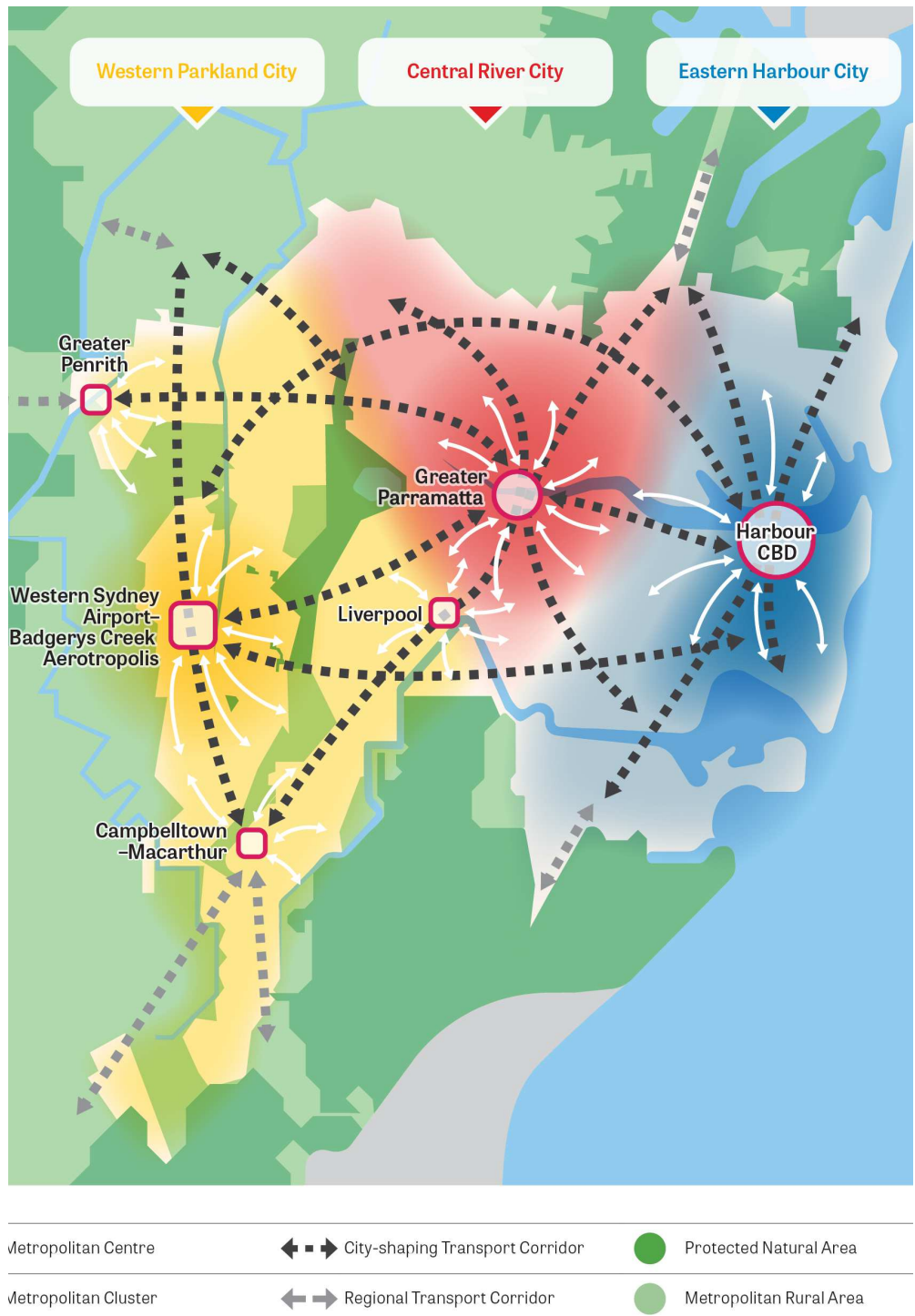}
    \caption{Schematic representation of Sydney's arterial traffic flows. Image provided by the Greater Sydney Commission with permission to reproduce from page 7 in~\cite{GreaterSydneyCommission2016}.}
    \label{fig:GreaterSydneyFlows}
\end{figure*}

\end{document}